\documentclass[twocolumn]{aastex631}
\usepackage{amsmath, mathtools}
\usepackage{courier}
\usepackage{url}

\shorttitle{}
\shortauthors{Claytor et al.}

\begin{document}

\title{Recovery of \textit{TESS} Stellar Rotation Periods Using Deep Learning}

\correspondingauthor{Zachary R. Claytor}
\email{zclaytor@hawaii.edu}

\author[0000-0002-9879-3904]{Zachary R. Claytor}
\affiliation{Institute for Astronomy, University of Hawai‘i at Mānoa, 2680 Woodlawn Drive, Honolulu, HI 96822, USA}

\author[0000-0002-4284-8638]{Jennifer L. van Saders}
\affiliation{Institute for Astronomy, University of Hawai‘i at Mānoa, 2680 Woodlawn Drive, Honolulu, HI 96822, USA}

\author[0000-0003-4450-0368]{Joe Llama}
\affiliation{Lowell Observatory, 1400 West Mars Hill Road, Flagstaff, AZ 86001, USA}

\author[0000-0002-7354-5461]{Peter Sadowski}
\affiliation{Department of Information and Computer Sciences, University of Hawai‘i at Mānoa, 1680 East-West Road, Honolulu, HI 96822, USA}

\author[0000-0001-9848-7483]{Brandon Quach}
\affiliation{Department of Computing and Mathematical Sciences, California Institute of Technology, 1200 E. California Blvd., MC 305-16
Pasadena, CA 91125, USA}

\author[0000-0003-1719-5046]{Ellis A. Avallone}
\affiliation{Institute for Astronomy, University of Hawai‘i at Mānoa, 2680 Woodlawn Drive, Honolulu, HI 96822, USA}

\submitjournal{AAS Journals 2021 April 28}

\begin{abstract}
    We used a convolutional neural network to infer stellar rotation periods from a set of synthetic light curves simulated with realistic spot evolution patterns. 
    We convolved these simulated light curves with real \textit{TESS} light curves containing minimal intrinsic astrophysical variability to allow the network to learn \textit{TESS} systematics and estimate rotation periods despite them.
    In addition to periods, we predict uncertainties via heteroskedastic regression to estimate the credibility of the period predictions. 
    In the most credible half of the test data, we recover 10\%-accurate periods for 46\% of the targets, and 20\%-accurate periods for 69\% of the targets. 
    Using our trained network, we successfully recover periods of real stars with literature rotation measurements, even past the 13.7-day limit generally encountered by \textit{TESS} rotation searches using conventional period-finding techniques.
    Our method also demonstrates resistance to half-period aliases.
    We present the neural network and simulated training data, and introduce the software \texttt{butterpy} used to synthesize the light curves using realistic star spot evolution.
\end{abstract}

\section{Introduction}
Stellar rotation is fundamentally linked to the structure and evolution of stars.
In the decade since \textit{Kepler}, much has been learned about rotation, feeding into asteroseismology, empowering gyrochronology, and changing the way we think about stellar evolution codes.
Rotation period estimates are made possible through a variety of methods.
Historically, spectroscopy enabled estimates of rotation velocity due to Doppler red/blue shift from the receding/approaching halves of the stellar disk.
The projected rotation velocity could then be used to compute an upper limit on the period if the stellar radius was known.
Missions like \textit{CoRoT} \citep{Baglin2006} and \textit{Kepler} \citep{Borucki2010} have shifted the paradigm: the majority of period estimates now employ photometry instead of spectroscopy.
This works particularly for stars which, like the Sun, exhibit magnetic dark and bright spots that induce periodic variations to the light curves as the stars rotate.
Several techniques have been developed in recent years to extract rotation information from spot-modulated stellar light curves.
Namely, Lomb-Scargle periodograms \citep{Marilli2007, Feiden2011}, autocorrelation analysis \citep{McQuillan2013, McQuillan2014}, wavelet transforms \citep{Mathur2010, Garcia2014}, Gaussian processes \citep{Angus2018}, and combinations of these \citep{Ceillier2017, Santos2019, Reinhold2020} have all been used to infer rotation periods from light curves.

Period-finding methods have paved the way for large studies of stellar rotation. Applied to \textit{CoRoT} and \textit{Kepler}, these techniques have delivered tens of thousands of rotation period estimates, which in turn have been used to advance our understanding of stellar and Galactic evolution \citep[e.g.,][]{McQuillan2014, vanSaders2016, vanSaders2019, Davenport2017, Claytor2020, Amard2020}.
The \textit{Transiting Exoplanet Survey Satellite} \citep[\textit{TESS},][]{Ricker2020} stands to increase the number of inferred periods by an order of magnitude in its ongoing all-sky survey.

Rotation studies have also brought to light the limitations of period detection methods.
For example, conventional methods are subject to aliases, and they still struggle to detect rotation in quiet, Sun-like stars \citep{McQuillan2014, vanSaders2019, Reinhold2020}.
Furthermore, the traditional methods do not necessarily reveal a star's true period. 
Rather, they reveal the period(s) of latitudes at which star spots form, which may rotate faster or slower than the star's equator due to surface differential rotation.

Finally, the systematics of \textit{TESS} have made traditional period searches difficult \citep{Oelkers2018a, CantoMartins2020, Holcomb2020, Avallone2021}.
The lunar-synchronous orbit of \textit{TESS} has a 13.7-day period, and the telescope is subject to background variations from reflected sunlight causing periodic contamination that is difficult to remove.
As a result, dedicated rotation studies struggle to obtain reliable periods longer than about 13 days \citep[e.g.,][]{CantoMartins2020, Holcomb2020, Avallone2021}.
New, data-driven methods are needed to overcome these systematics and recover periods.

Deep Learning is relatively new to astronomy, but in a short time deep learning methods have proven to be valuable at mining information from large data sets.
Neural networks, and in particular Convolutional Neural Networks (CNNs), efficiently extract information from time series, spectra, and image data.
The strength of CNNs comes from their local connectivity, which incorporates the knowledge that neighboring input points are highly correlated.
Examples of success using CNNs in astronomy include \citet{Hezaveh2017}, who used CNNs to characterize gravitational lenses from image data.
Within the realm of stellar astrophysics, \citet{Guiglion2020} used the same techniques to obtain stellar parameters from spectra.
Moreover, \citet{Feinstein2020} and \citet{Blancato2020} used CNNs to infer stellar parameters and flare statistics from light curves.

Using convolutional neural networks, we predict stellar rotation periods from wavelet transforms of light curves.
The use of supervised machine learning requires the existence of a training data set for which the target, in this case the rotation period, is known.
This is not yet possible with \textit{TESS} due to the difficulties of obtaining reliable periods using traditional techniques.
Furthermore, while there is some overlap between \textit{TESS} and \textit{Kepler}, the \textit{TESS} observations of the \textit{Kepler} field are short, spanning only 27 days.
This is enough to recover and validate short rotation periods \citep{Blancato2020}, and possibly some subset of longer periods \citep{Lu2020}, but not enough to be useful for the broader population of stars in our Galaxy.
Moreover, even large rotation samples from \textit{Kepler} are likely contaminated with mismeasured periods \citep{Aigrain2015}.
Using a training set of periods obtained with conventional techniques risks imprinting this contamination onto the neural network.
To avoid this, we followed the approach of \citet{Aigrain2015} and used a set of synthetic light curves generated from physically motivated star spot emergence models.
This is an example of simulation-based inference \citep[e.g.,][]{Cranmer2020}, wherein we simulate a physical process and use machine learning to address the inverse problem of inferring the rotation.

We introduce \texttt{butterpy}\footnote{\url{https://github.com/zclaytor/butterpy} \citep{butterpy}.}, an open-source Python package designed to simulate realistic star spot emergence and synthesize light curves, followed by a description of the input physics of the simulations.
After describing our training set, we outline our convolutional neural network and the methods we use to train, validate, and test the network.
We evaluate our trained neural network on synthesized data sets spanning different period ranges to identify for what periods the network is most predictive.
Next, we discuss the network's performance on a small set of real light curves for which rotation periods are known.
We also compare our network predictions to periods recovered using conventional methods before finally concluding with thoughts on the feasibility of our methods to recover stellar rotation periods from real \textit{TESS} light curves.

\section{Synthetic Light Curves: \texttt{butterpy}}
Synthesized light curves have several advantages over observed light curves: (1) the true, equatorial period of the simulated star is known, rather than an estimate of the period (which may be wrong), (2) data of any length and cadence can be synthesized, and (3) other physical properties like spot characteristics, differential rotation, and surface activity are known and can be independently probed. 

To simulate light curves, we developed \texttt{butterpy}, a Python package designed to generate realistic, physically motivated spot emergence patterns faster than conventional surface flux transport codes.
The name \texttt{butterpy} comes from the butterfly-shaped pattern of spot emergence with time exhibited by the Sun \citep[e.g., ][]{Hathaway2015}.
We built upon the software developed by \citet{Aigrain2015}, which relied on the flux transport models of \citet{Mackay2004} and \citet{Llama2012} to generate spot emergence distributions which were in turn used to compute light curves.
The original model of \citet{Mackay2004} was designed to reproduce spot emergence patterns of the Sun as well as Zeeman Doppler images of the pre-main-sequence star AB Doradus.
Later, \citet{Llama2012} used this model in tandem with exoplanet transit observations to trace the migration of active latitude bands across the surfaces of stars.
\citet{Aigrain2015} used the light curves generated from these spot distributions to test the recovery rates of various period detection techniques, which we seek to emulate. We discuss the method and assumptions here for clarity. 

\subsection{Spot Emergence and Light Curve Computation}
There are several variables to consider regarding the emergence of star spots and their effect on the star's light curve, including the latitudes and rates of emergence, the spot lifetimes, and the rotation speed at the latitude of emergence if the star rotates differentially.
While observations of these on stars other than the Sun are limited, they are very well characterized for the Sun \citep[][and references therein]{Hathaway2015}.
For our model, we therefore start with the characteristics that are known for the Sun and allow them to vary.

\subsubsection{Location and rate of spot emergence}
The latitudes of spot emergence on the Sun vary with the Sun's 11-year activity cycle.
At the beginning of a cycle, spots emerge within active regions at high latitudes \citep[$\lambda \approx \pm 30^\circ$,][]{Hathaway2015}, and the latitude of emergence migrates toward the equator throughout the rest of the cycle.
Before the cycle ends, new spot groups begin forming again at high latitudes, indicating some amount of overlap between consecutive cycles.
The repeating decay in spot latitude with time gives rise to a butterfly-like pattern known as a ``butterfly diagram".
Butterfly patterns have been observed in other stars as well \citep{Bazot2018, Nielsen2019, Netto2020}, but some stars show a random distribution of spot emergence latitudes with time \citep[e.g.,][]{Mackay2004}.
The width of active latitude bands has also been shown to differ even for Sun-like stars \citep{Thomas2019}.
In our model, we allow for either a random or butterfly-like spot emergence between a minimum and maximum latitude that are unique for each star.
For the butterfly pattern, spots begin the cycle emerging at a latitude $\lambda_\mathrm{max}$, decaying exponentially with time to latitude $\lambda_\mathrm{min}$ at the end of the cycle \citep{Hathaway2011}. 

As for longitude, spots on the Sun tend to emerge in groups, either next to existing spots, or in some cases antipolar to existing spots \citep{Hathaway2015}.
Less often, spots will emerge at random longitudes, not necessarily associated with any existing spot groups.
We respectively refer to these two cases as correlated and uncorrelated active regions. 
In our simulations, we follow the approach of \citet{Aigrain2015}, dividing the stellar surface into 16 latitude and 36 longitude bins; the probability of spot emergence is distributed across these bins.
To account for the relative likelihood of correlated and uncorrelated emergence, bins already containing active regions are assigned a higher probability of emergence.

The rates of sunspot emergence change with spot area and with time throughout the activity cycle.
\citet{Schrijver1994} expressed the number of spots emerging in area interval ($a$, $a + \mathrm{d}a$) and time interval ($t$, $t + \mathrm{d}t$) as $r(t) a^{-2}~\mathrm{d}a~\mathrm{d}t$, where $r(t)$ represents the time-varying emergence rate amplitude, the active region area $a$ is in square degrees, and $t$ is the time elapsed in the activity cycle, ranging from zero to one.
For the time dependence, spots emerge very slowly at the beginning, more rapidly in the middle, and slowly again at the end \citep{Hathaway1994}.
\citet{Mackay2004} modeled this using a squared sine function: $r(t) = A \sin^2(\pi t)$.
The activity level $A$ is an adjustable scale factor controlling both the average rate of spot emergence and the amplitude of light curve modulation of a single spot.
It is defined such that $A = 1$ for the Sun.

\subsubsection{Latitudinal Differential Rotation}
We define heliographic longitude such that $\phi = 0$ always faces the Earth.
As a consequence, spots move in longitude as the star rotates.
The Sun rotates more rapidly near the equator than at the poles, a phenomenon known as ``latitudinal differential rotation" (henceforth just ``differential rotation").
While differential rotation is more difficult to observe on other stars, some stars may exhibit ``anti-solar" differential rotation, wherein the equator rotates more slowly than the poles.
This has been observed particularly in slowly-rotating stars \citep[e.g.,][]{Rudiger2019}.
In our model, we allow for solar-like, anti-solar, and solid-body rotation.
Following \citet{Aigrain2015}, we model the differential rotation profile as
\begin{equation}
    \begin{aligned}
        \phi_k(t) &= \phi_k(t_{\mathrm{max},k}) + \Omega(\lambda_k) (t - t_{\mathrm{max},k}),\\[10pt]
        \Omega(\lambda_k) &= \frac{2\pi}{P_\mathrm{eq}}(1 + \alpha \sin^2 \lambda_k).
    \end{aligned}
\end{equation}
Here, $\lambda_k$ and $\phi_k$ denote the heliographic latitude and longitude of spot $k$.
$t_{\mathrm{max},k}$ is the time at which the spot achieves maximum flux, $\Omega$ is the angular velocity at latitude $\lambda_k$, and $\alpha$ is the differential rotation shear parameter.
To include anti-solar, solid-body, and solar-like profiles, we allow $\alpha$ to range from -1 to 1.

\subsubsection{Spot-Induced Flux Modulation}
Once spot emergence is determined, we simulate spot evolution and flux modulation based on the simplified model of \citet{Aigrain2012, Aigrain2015}.
They take the photometric signature $\delta F_k(t)$ of a single spot $k$ to be
\begin{equation}
    \begin{aligned}
        &\delta F_k(t) = f_k (t)~\max\{\cos\beta_k(t),~0\}\label{flux},\\
        &\cos\beta_k(t) = \cos\phi_k(t) \cos\lambda_k \sin i + \sin\lambda_k \cos i,
    \end{aligned}
\end{equation}
where $\beta_k(t)$ is angle between the spot normal and the line of sight, accounting for projection on the stellar surface.
The inclination $i$ is the angle between the rotation axis and the line of sight, and $\lambda_k$ and $\phi_k(t)$ are again the latitude and longitude of the spot.
The factor $f_k$ is the amount of luminous flux removed if spot $k$ is observed at the center of the stellar disk.
\citet{Aigrain2015} used an exponential rise and decay to model the rapid rise and slow decay of single spots, but we employ a two-sided Gaussian to exhibit smoothness while preserving the same emergence and decay behavior:
\begin{equation}
    \begin{aligned}
        &f_k(t) = f_k^\mathrm{max} \exp\left[-\left(t - t_{\mathrm{max}, k}\right)^2/\tau^2\right],\\[10pt]
        &\tau = 
        \begin{dcases}
            \tau_\mathrm{emerge} = \max\left\{2~\mathrm{d},\ \frac{\tau_\mathrm{spot}}{5} P_\mathrm{eq}\right\},~t \leq t_{\mathrm{max}, k} \\[5pt]
            \tau_\mathrm{decay} = \tau_\mathrm{spot} P_\mathrm{eq},~t > t_{\mathrm{max}, k},\\[5pt]
        \end{dcases}
    \end{aligned}
\end{equation}
where $f_k^\mathrm{max}$ is the flux removed by spot $k$ at the time of maximum emergence $t_{\mathrm{max}, k}$, $\tau$ is the relevant emergence or decay timescale, and $\tau_\mathrm{spot}$ is a dimensionless parameter used to relate the emergence and decay timescales to the equatorial rotation period $P_\mathrm{eq}$.
Like \citet{Aigrain2015}, we parametrized the emergence and decay timescales as multiples of the equatorial rotation period.
The form of the emergence timescale was chosen so that, in general, emergence is five times faster than decay, with a minimum possible emergence timescale of two days.
In the simple model of \citet{Aigrain2012}, $f_k^\mathrm{max}$ takes into account the spot area and contrast, but the model of \citet{Aigrain2015} relates this factor to the strength of the magnetic field:
\begin{equation} \label{fk}
    f_k^\mathrm{max} = 3 \times 10^{-4}\ A\ B_{r,k}/\langle B_{r,k}\rangle_k,
\end{equation}
where $A$ is the activity level, and the constant is chosen such that $A = 1$ reproduces approximately Sun-like behavior.

With this expression, the single-spot luminous flux modulation is proportional to the strength of the radial magnetic field at that spot.
The magnetic field strength or magnetic flux is proportional to the area of the active region, which \citet{vanBallegooijen1998} derive using the angular width of magnetic bipoles emerging from the active region:
\begin{equation} \label{magnets}
    B^{\pm}_r (\theta,~\phi) = B_\mathrm{max} \left( \frac{\beta_\mathrm{init}}{\beta_0}\right)^2
    \exp\left[-\frac{2(1-\cos\beta_{\pm}(\theta,~\phi))}{\beta_0^2}\right],
\end{equation}
where $B^{\pm}_r$ is the radial component of the magnetic field near either the positive or negative bipole, $B_\mathrm{max}$ is the initial peak magnetic field strength in the active region, $\beta_\mathrm{init}$ is the angular width of a single bipole, $\beta_0$ is the angular width (in degrees) of the bipole at the time the active region is inserted into the model, accounting for diffusion, and $\beta_{\pm}$ is the heliocentric angle between a field point and one of the bipoles.

\citet{vanBallegooijen1998} assumed that the bipole width $\beta_\mathrm{init}$ is proportional to the angular separation between the positive and negative poles, which they call $\Delta \beta$, with a proportionality factor of 0.4.
Assuming spots form within ten degrees of the active region bipoles, the value of the exponential factor differs from unity by less than one percent.
For this reason, we approximate the exponential factor as unity.
Thus, at the location of a star spot, Equation~(\ref{magnets}) simplifies to
\begin{equation}
    B_r \approx B_\mathrm{max} \left( \frac{0.4 \Delta \beta}{\beta_0}\right)^2.
\end{equation}
Combining this with Equation~(\ref{fk}),
\begin{equation}
    f_k^\mathrm{max} = 3 \times 10^{-4} A \left( \frac{\Delta \beta_k}{\langle \Delta \beta_k\rangle_k}\right)^2.
\end{equation}

\citet{vanBallegooijen1998} and \citet{Mackay2004} consider a range of bipole widths from about 3.5$^\circ$ to 10$^\circ$.
The distribution of $\Delta \beta$ is the same for every star, so $\langle \Delta \beta_k \rangle_k$ is effectively constant.
Putting it all together, we have a final system of equations to describe the change in luminous flux from a single spot:
\begin{equation}
    \begin{aligned}
        &\delta F_k(t) = f_k(t) \exp\left[-\left(t - t_{\mathrm{max}, k}\right)^2/\tau^2\right], \\[10pt]
        &f_k(t) = 3 \times 10^{-4} A \left( \frac{\Delta \beta_k}{\langle \Delta \beta_k\rangle_k}\right)^2~\max\{\cos\beta_k(t),~0\}, \\[10pt]
        &\tau = 
        \begin{dcases}
            \max\left\{2~\mathrm{d},\ \frac{\tau_\mathrm{spot}}{5} P_\mathrm{eq}\right\},\ t \leq t_{\mathrm{max}, k} \\
            \tau_\mathrm{spot} P_\mathrm{eq},\ t > t_{\mathrm{max}, k}\\
        \end{dcases},\\[10pt]
        &\cos\beta_k(t) = \cos\phi_k(t) \cos\lambda_k \sin i + \sin\lambda_k \cos i.
    \end{aligned}
\end{equation}
This depends on $\Delta \beta_k$, $\lambda_k$, $\phi_k(t)$, and $t_{\mathrm{max},k}$, which are unique to each spot; and $A$, $\tau_\mathrm{spot}$, $P_\mathrm{eq}$, and $i$, which are unique to each star.

\subsection{Training Set}
\begin{figure}
    \centering
    \includegraphics[width=\linewidth]{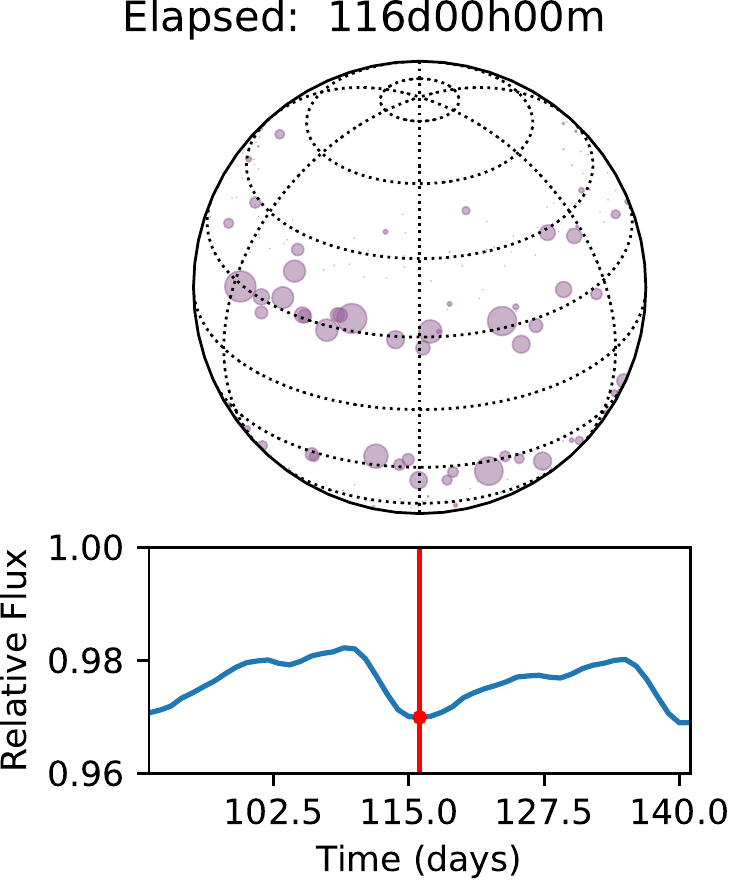}
    \caption{An example of a \texttt{butterpy} simulation of spot evolution and light curve generation. This figure is available online as an interactive figure. The online figure has an interactive slider and play/pause buttons that allow the user to move the figure through time and see changes in the light curve as spots rotate into and out of view.}
    \label{fig:surface}
\end{figure}

\begin{deluxetable*}{lll}
\tablecaption{Distribution of Simulation Input Parameters}
\tablehead{\colhead{Parameter} & \colhead{Range} & \colhead{Distribution}}
\startdata
Equatorial rotation period $P_\mathrm{eq}$ & 0.1 -- 180 days & uniform \\
Activity level $A$ & 0.1 -- 10 $\times$ solar & log-uniform \\
Activity cycle length $T_\mathrm{cycle}$ & 1 -- 40 years & log-uniform \\
Activity cycle overlap $T_\mathrm{overlap}$ & 0.1 year -- $T_\mathrm{cycle}$ & log-uniform \\ 
Minimum spot latitutde $\lambda_\mathrm{min}$ & 0$^\circ$ -- 40$^\circ$ & uniform \\
Maximum spot latitude $\lambda_\mathrm{max}$ & $\lambda_\mathrm{min}+5^\circ$ -- 80$^\circ$ & uniform \\
Spot lifetime $\tau_\mathrm{spot}$ & 1 -- 10 & log-uniform \\
Inclination $i$ & 0$^\circ$ -- 90$^\circ$ & uniform in $\sin^2 i$ \\
Latitudinal rotation shear $\Delta \Omega / \Omega_\mathrm{eq}$ & 0.1 -- 1 (50\%) & log-uniform \\
 & 0 (25\%) & \\
 & -1 -- -0.1 (25\%) & log-uniform \\
\enddata
\tablecomments{We adopted the distributions used by \citet{Aigrain2015} with minor modifications: (1) we sampled a broader range of periods and activity levels, (2) we used a uniform distribution of periods so as not to impart unwanted bias on the neural network prediction, (3) we include anti-solar differential rotation by allowing the shear parameter to be negative.}
\label{tab:simulations}
\end{deluxetable*}

Using the model in \texttt{butterpy}, we generated one million light curves at thirty-minute cadence and one-year duration to match the \textit{TESS} Full-Frame Images (FFIs) in the continuous viewing zones (CVZs).
The simulation input parameters are listed in Table~\ref{tab:simulations}.
We sampled periods uniformly from the range [0.1, 180] days.
The period range was chosen to be as wide as possible to simulate the fastest-rotating stars ($P_\mathrm{eq} \approx 0.1$ day) while also capturing anything that would go through at least two rotations under observation in the \textit{TESS} CVZs (the total baseline is $\sim$360 days, so an object with $P_\mathrm{eq} = 180$ will go through exactly two rotations in that time).
We chose the remaining distributions and ranges to reflect those of \citet{Aigrain2015}, with minor adjustments in the ranges of activity level and differential rotation shear to search a broader parameter space.
Our input distributions assumed no relation between rotation period and activity level.
Figure~\ref{fig:surface} illustrates an example simulation, showing the distribution of spots on the surface as well as their impact on the observed light curve.

\subsection{\textit{TESS} Noise Model}
To ensure the training light curves properly emulate real \textit{TESS} light curves, the training set must exhibit \textit{TESS}-like noise.
\citet{Aigrain2015} used light curves from quiescent \textit{Kepler} stars to achieve this.
In their study, a sample of stars from \citet{McQuillan2014} with no significant period detection served as the quiescent data set.
Because there are no existing bulk period measurements for stars in the CVZs, we must find another means of simulating \textit{TESS} noise.

While \textit{TESS} is a planet-finding mission, the southern CVZ contains thousands of galaxies which should have roughly constant brightness with time.
Any changes in the light curves of these galaxies would be due solely to \textit{TESS} instrument systematics.
Thus, the galaxy light curves should reasonably resemble light curves of quiescent stars in \textit{TESS}. 

We selected roughly 2,000 galaxies in the southern CVZ with $Tmag \leq 15$ as our quiescent sample, removing a handful of galaxies known to be active and in the Half-Million Quasars catalog \citep{Flesch2015}.
We queried FFI cutouts from the Mikulski Archive for Space Telescopes (MAST) using \texttt{Lightkurve} and \texttt{TESScut} \citep{Lightkurve2018, Brasseur2019}.
Then, we performed background subtraction and aperture photometry on each source using \texttt{Lightkurve} regression correctors, following \citet{LightkurveDocs}.
To summarize, aperture masks were chosen using the \texttt{create\_threshold\_mask} function in \texttt{Lightkurve}.
This method selects pixels with fluxes brighter than a specified threshold number of standard deviations above the image's median flux value. 
We specified thresholds based on the target's brightness to exclude background pixels from the aperture.
Once the raw light curve was computed, the regression correctors fit principle components of the time-series images and subtracted the strongest components from the raw light curve.
All sector light curves for a source were then median-normalized and stitched together to form the final ``pure noise" light curve.

The galaxy light curves were linearly interpolated to each \textit{TESS} cadence to fill gaps, whether for missing observations or entire missing sectors.
Cadences missing at the beginning or end of the light curve were filled with the light curve's mean flux value.
Finally, a galaxy light curve was chosen at random to be convolved with each of the synthetic light curves, yielding our final set of simulated \textit{TESS}-like light curves.
We divided the set of 2,000 galaxies into two sets of 1,000: one set to be convolved with light curves from the training partition, and one for the validation and test partitions (see Section~\ref{sec:cnn} for more about data partitioning).

\section{Data Processing/Wavelet Transform}
\begin{figure}
    \centering
    \includegraphics[width=\linewidth]{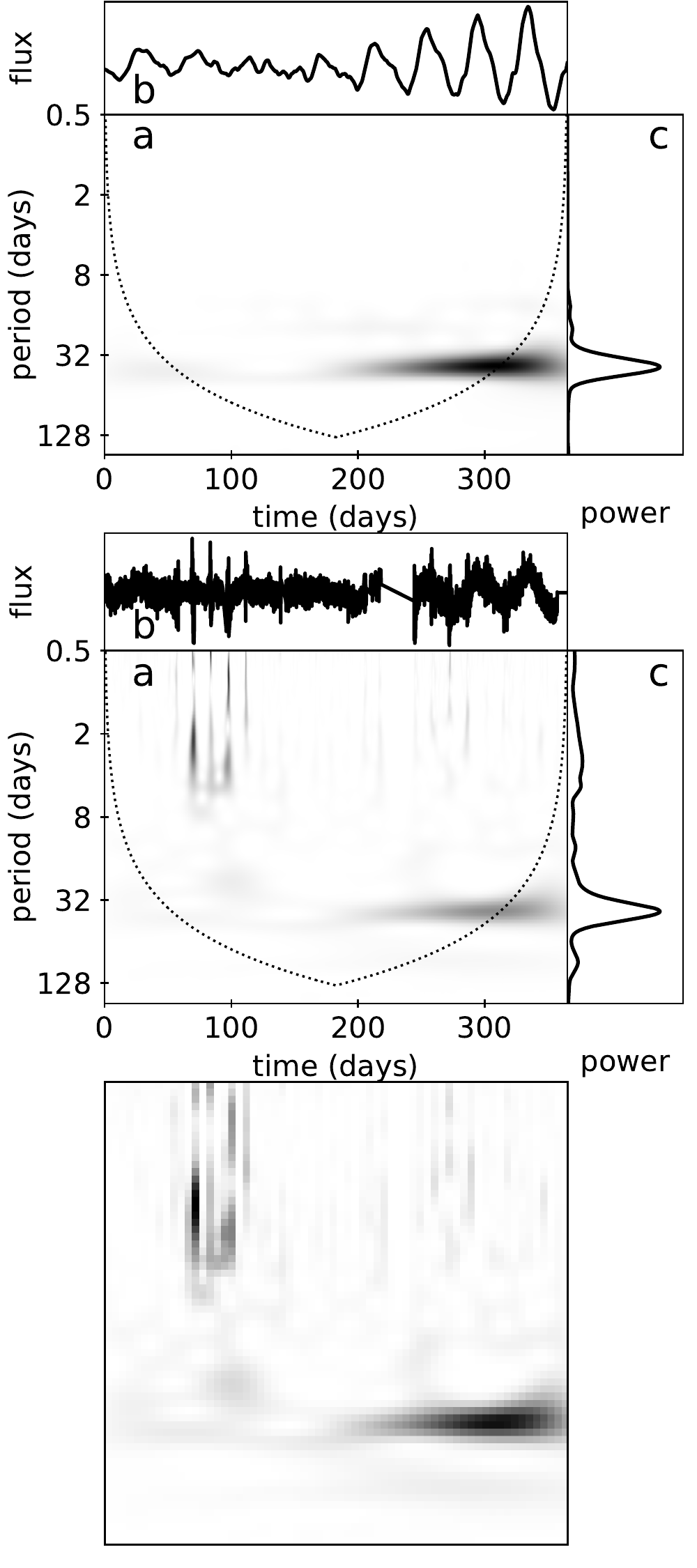}
    \caption{
        \textit{Top:} Morlet wavelet transform (a) of a noiseless light curve, shown with the light curve (b) and global wavelet power spectrum (c).
        \textit{Center:} plots for the same light curve convolved with \textit{TESS} noise and re-stitched.
        The dotted curve marks the cone of influence, below which the power spectrum is susceptible to edge effects.
        \textit{Bottom:} example of a binned wavelet power spectrum we used to train our neural network.
        Neural networks can learn to ignore the noise and pick out stellar signals.}
    \label{fig:wps}
\end{figure}

\begin{deluxetable*}{lccccc}
\tablecaption{Convolutional Neural Network Architecture}
\tablehead{\colhead{Layer Type} & \colhead{Number of Filters} &\colhead{Filter Size} & \colhead{Stride} & \colhead{Activation} & \colhead{Output Size} \\} 
\startdata
Input image &  -                & -             & -         &               & $64\times64$ \\
Conv2D      & 8                 & $3\times3$    & $1\times1$& ReLU          & $62\times62\times8$\\
MaxPool2D   & 1                 & $1\times3$    & $1\times3$& -             & $62\times20\times8$\\
Conv2D      & 16                & $3\times3$    & $1\times1$& ReLU          & $60\times18\times16$\\
MaxPool2D   & 1                 & $1\times3$    & $1\times3$& -             & $60\times6\times16$\\
Conv2D      & 32                & $3\times3$    & $1\times1$& ReLU          & $58\times4\times32$\\
MaxPool2D   & 1                 & $1\times4$    & $1\times4$& -             & $58\times1\times32$\\
Flatten     & -                 & -             & -         & -             & 1856 \\
Dense       & -                 & -             & -         & ReLU          & 256 \\
Dense       & -                 & -             & -         & ReLU          & 64 \\
Dense       & -                 & -             & -         & Softplus      & 2 \\
\enddata
\label{tab:architecture}
\tablecomments{We use three 2D convolution layers, each with ReLU activation and max-pooling. Our implementation uses 2D max-pooling with a 1-dimensional kernel to achieve pooling in the time dimension but not the frequency dimension. This choice preserves frequency resolution but achieves a small amount of translational invariance in the time dimension. The output of the convolution block is flattened to a 1-dimensional array and passed through three fully connected (dense) layers, with ReLU and finally softplus output, to yield two numbers: the rotation period and its uncertainty.}
\end{deluxetable*}

There are several options for input to a neural network to predict rotation periods.
One could use the light curve directly; \citet{Blancato2020} suggest this as the best way to obtain periods using neural networks without loss of information.
However, using the light curve as input means that the information conveying periodicity is temporally spread out.
While neural networks can certainly learn to predict periods this way, a frequency representation concentrates the period information to one location in input space.
Lomb-Scargle periodograms \citep{Lomb1976, Scargle1982, Feiden2011} and autocorrelation functions \citep{McQuillan2013, McQuillan2014} are two tried-and-true methods of period estimation that have some promise as input to neural networks.
While these methods are effective at concentrating periodicity information to one location, real stars' observed periodicity can change with time due to differential rotation.
Lomb-Scargle and autocorrelation methods average over these changes, potentially blurring out interesting evolution.
The continuous wavelet transform \citep{Torrence1998} has also been used to identify rotation periods from stellar light curves \citep{Mathur2010, Garcia2014, Santos2019} and it has the bonus of elucidating changes in periodicity with time.
We chose this method to localize periodic information while allowing the tracing of spot evolution. 

We used the continuous wavelet transform implemented in \texttt{SciPy} \citep{Scipy2020} with the power spectral density correction of \citet{Liu2007}.
Using a Morlet wavelet, we computed wavelet power spectra for both the noiseless and noise-injected light curves in our training set.
Examples of both noiseless and noisy power spectra are shown for the same simulated star in Figure~\ref{fig:wps}. 
We then rebinned the power spectra to $64\times64$ pixels and saved them as arrays for fast access.
Larger binned sizes were tested (e.g., $128\times128$) and showed no significant change in performance.

We ran several tests with the period axis of the wavelet power spectra, trying maximum periods of 128, 150, and 180 days, before settling on 180 days (i.e., half the observing window) for the final data products.
In the tests with 128 and 150 days, the neural network had some success predicting periods longer than the maximum visible period in the periodogram, even for the noisy data.
This suggests that neural networks can predict periods even when the period at maximum power is beyond the range of the plot, consistent with the results of \citet{Lu2020}.
This is encouraging for period predictions for stars outside the \textit{TESS} continuous viewing zones, where observations are substantially less than a year in duration.
In the end, we chose 180 days as the maximum value on the period axis to preserve the strongest rotation signals in as many objects as possible.

In addition to \texttt{butterpy}, our final data products will be made publicly available and include (1) the noiseless, synthesized light curves, (2) the normalized \textit{TESS} galaxy light curves, and (3) the binned wavelet transform arrays for both the noiseless and noisy light curves.

\section{Convolutional Neural Network}
\label{sec:cnn}
We used a convolutional neural network to predict rotation periods from wavelet transforms.
Table~\ref{tab:architecture} outlines the CNN architecture.
We used a sequence of 2D convolution layers with rectified linear activation (``rectifier" or ``ReLU") followed by 1D max-pooling in the time dimension.
The ReLU activation function has the form $f(x) = \max(0,~x)$.
Its nonlinearity allows the model to represent complex functions, and ReLU learns faster than other nonlinear activation functions.
Max-pooling is used to down-sample input and impart a small amount of translational invariance.
The shapes of the convolution and pooling kernels were chosen to impart equivariance in the frequency dimension (no pooling, since frequency is what we want to estimate) and translational invariance in the time dimension.
This means that the CNN will treat periodic signals the same regardless of when they occur in the wavelet power spectrum \citep[see Ch.~9 of][]{Goodfellow2016}.

The output of the convolution layers is then flattened to one dimension and fed into a series of three fully connected layers, also with ReLU activation.
The final layer uses softplus activation, which has the form $f(x) = \ln(1 + e^x)$. A smooth approximation to the rectifier, softplus activation ensures positive output while preserving differentiability.
The final layer outputs two numbers, which represent the rotation period and the period uncertainty.

The 2D wavelet power spectra were used as input to our neural network, while the corresponding model rotation periods served as the target output. 
Target periods were min-max scaled over the entire data set to the range [0, 1]. 
Each power spectrum array was min-max scaled to the range [0, 1] separately---using the min and max over the entire data set suppressed lower-amplitude signals and substantially impaired performance. 

Our full data set of one million examples was partitioned into three sets for model training, validation, and testing.
The training set consisted of 80\% and was used to fit the model weights.
The validation set (10\%) was used for early stopping (we stop training when the validation loss does not improve over a window of 10 training epochs to avoid overfitting) and for choosing the optimal hyperparameters.
The test set (10\%) was used for final model evaluation.

We used the Adam optimizer \citep{Kingma2014}, which allows for a variable learning rate, with negative log-Laplacian likelihood as the loss function.
This loss function allows us to predict both the rotation period and its error (a process known as heteroskedastic regression), indicating which period predictions are more reliable than others.
It has the form
\begin{equation}
    \mathcal{L} = \ln\left(2b\right) + \frac{|P_\mathrm{true} - P_\mathrm{pred}|}{b},
\end{equation}
where $b$ is taken to represent the predicted uncertainty.

Maximizing the log-likelihood of the Laplace distribution is equivalent to minimizing the mean absolute error instead of the mean-squared error, or predicting the median period instead of the mean under uncertainty.
This also means that in cases where the neural network cannot predict with high confidence, predictions will be biased toward the median of the period range.
Formally, the Laplace distribution has variance $2b^2$ and standard deviation $\sqrt{2}b$, but we use $b$ to represent the uncertainty for simplicity since we use it only to determine the relative credence of predictions.
Thus, our predicted uncertainties should not be considered statistically formal.

With 800,000 input-output pairs in the training set, our model takes roughly 3 hours until fully trained on a single NVIDIA RTX2080.
Once trained, evaluation on the test input of 100,000 wavelet power spectrum plots takes less than a minute.

\section{Results}
We trained and evaluated the neural network on year-long simulations of both noiseless and noise-injected wavelet transform images. 
We additionally used Lomb-Scargle periodograms, autocorrelation functions, and wavelet transforms to obtain independent period estimates from the noisy data.

\citet{Aigrain2015} performed blind injection-recovery exercises on synthesized \textit{Kepler}-like light curves to assess the reliability of conventional period-detection methods.
On average, the teams recovered periods with 10\% accuracy in $\sim$70\% of cases in which periods were obtained.
We adopt this 10\% accuracy threshold as our success metric, which we designate ``acc10."
In addition to acc10, we also quantify results with ``acc20," mean absolute percentage error (MAPE), and median absolute percentage error (MedAPE), defined as follows. 
If we define the absolute percentage error of example $i$ to be $\epsilon_i = |P_{\mathrm{pred},i} - P_{\mathrm{true},i}|/P_{\mathrm{true},i}$, then our recovery metrics are
\begin{equation}
    \begin{aligned}
    &\mathrm{MAPE} = \frac{1}{N}\sum_{i}^{N}\epsilon_i\\
    &\mathrm{MedAPE} = \mathrm{median}\{\epsilon_i\} \\
    &\mathrm{acc10} = \frac{1}{N}\sum_{i}^{N}H(0.1 - \epsilon_i)\\
    &\mathrm{acc20} = \frac{1}{N}\sum_{i}^{N}H(0.2 - \epsilon_i),\\
    \end{aligned}
\end{equation}
where $H(x - \epsilon_i)$ is the Heaviside or unit step function.

Before commenting on our period recovery, it is important to note the differences in our light curve sample from that of \citet{Aigrain2015}.
The most important differences are in the range of activity level and the light curve pre-processing.
Our sample spans a wider range of activity levels, ranging from 0.1 to 10 times Solar, as opposed to 0.3 to 3 times Solar in \citet{Aigrain2015}.
The logarithmic scale of the distribution ensures that the increase in range evenly adds examples to the high- and low-activity ends.
Thus, despite having higher-amplitude examples in our sample, there should be enough lower-amplitude examples to compensate, preserving the comparability of our summary statistics to those of \citet{Aigrain2015}.
Light curve pre-processing differs because of the differences in the \textit{Kepler} pipeline and our custom \textit{TESS} FFI pipeline.
In principle, the \textit{Kepler} pipeline more aggressively removes systematics, so the \citet{Aigrain2015} simulated light curves are cleaner than ours.

\subsection{Period recovery using conventional methods}

\begin{figure}
    \centering
    \includegraphics[width=\linewidth]{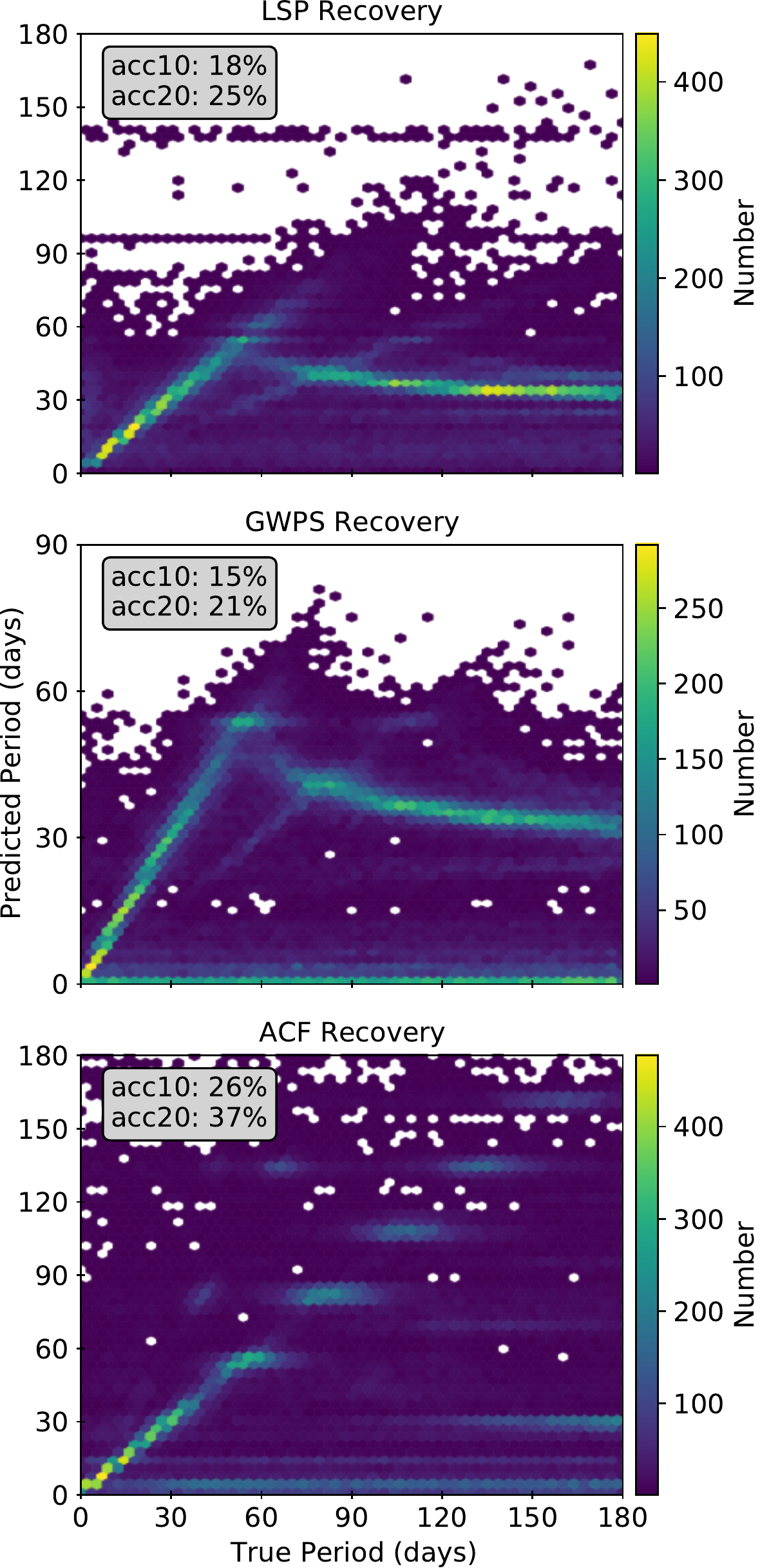}
    \caption{Period recovery using Lomb-Scargle periodogram (LSP, top), global wavelet power spectrum (GWPS, middle), and autocorrelation function (ACF, bottom). ``acc10" represents fraction of periods recovered to within 10\% accuracy, while ``acc20" is the recovery to within 20\%. ACF has the highest overall success, but the recovery worsens significantly at periods longer than about 30 days.}
    \label{fig:conv_tests}
\end{figure}

We recovered periods from our sample of noise-injected light curves using Lomb-Scargle periodograms \citep[LSP, as implemented in \texttt{Lightkurve},][]{Lightkurve2018}, autocorrelation functions (ACF, \citealp{McQuillan2013}, and as implemented in \texttt{starspot}, \citealp{Angus2018}), and global wavelet power spectrum \citep[GWPS, as implemented in \texttt{SciPy},][]{Scipy2020}.
The recovery results are summarized in Figure~\ref{fig:conv_tests}.
In each panel, objects falling within 10\% of the line $y = x$ are successfully recovered according to our metric.
All three methods struggle to recover periods longer than about 50 days.
Longer than this, LSP and GWPS mistakenly recover signals approaching 30 days in period, which we suspect represents the \textit{TESS} sector length of 27 days.
LSP and GWPS are also susceptible to half-period aliases, which fall along the line $y = \frac{1}{2} x$.
ACF is less susceptible to half-period aliases and is the most successful method overall.
However, the ACF and GWPS often misidentify signals at 5, 13, and 27 days (all well-known frequencies associated with \textit{TESS} telescope systematics) as the rotation period.
Interestingly, the ACF has small pockets of higher success at integer multiples of 27 days, beginning at 54 days.
These occur when a star rotates an integer number of times in an integer number of sectors.
The sector-to-sector stitching affects subsequent revolutions the same way, so the signal is preserved enough for the autocorrelation function to detect.

In general, recovery was better for targets with higher light curve amplitude for all three methods, as one would expect.
The recovery rates also improve when limited to shorter periods.
We have assumed no rotation-activity relation, so the improved recovery at shorter periods occurs when more rotations are observed in the given baseline, resulting in higher power in the periodograms.
Moreover, at shorter periods rotation signals are less easily lost in the telescope systematics.
If we limit to periods between 0 and 50 days, acc10 and acc20 improve to 43\% and 59\% for LSP, 43\% and 56\% for GWPS, and 36\% and 47\% for ACF.
Thus, for short periods, LSP achieved the highest rate of success.

\subsection{CNN performance on noiseless data}
\begin{figure*}
    \centering
    \includegraphics[width=\linewidth]{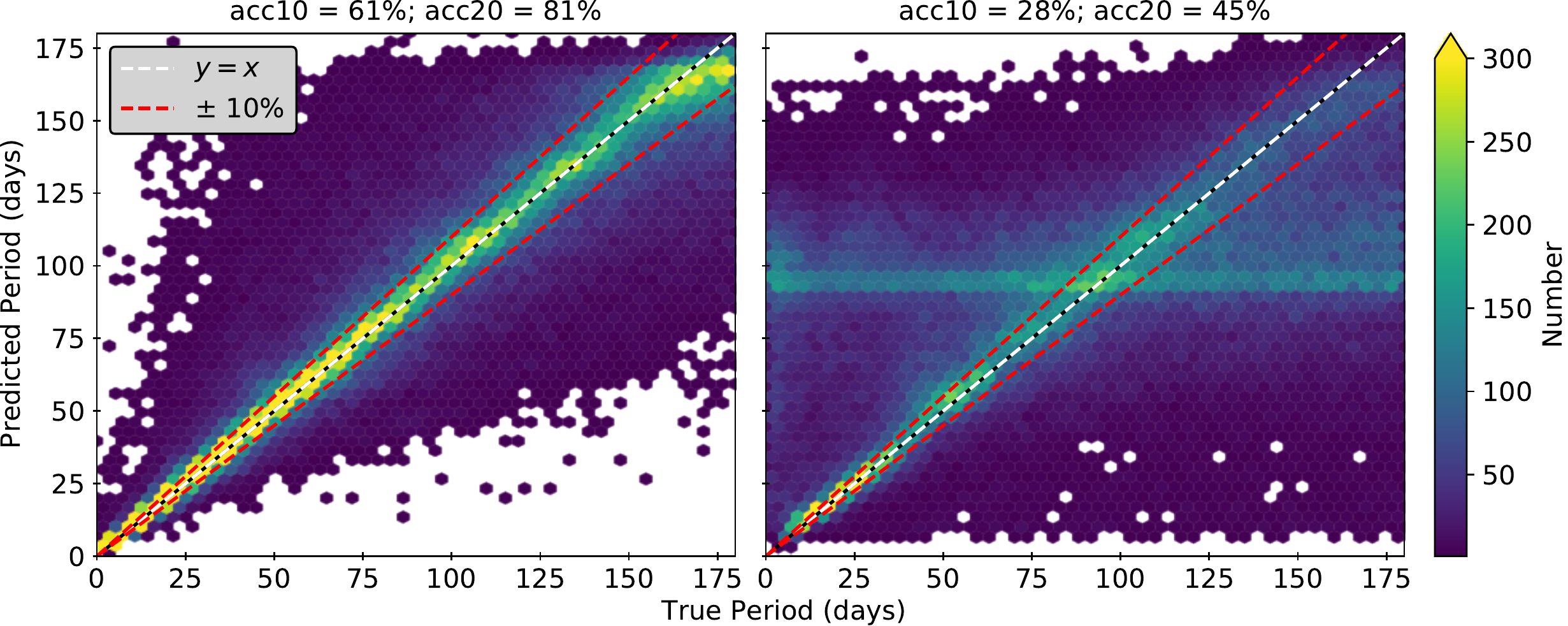}
    \caption{
        \textit{Left:} Period predictions by our convolutional neural network trained on wavelet transforms of noiseless light curves. Predicted periods have mean absolute percentage error of 14\%, median absolute percentage error of 7\%, acc10 of 61\%, and acc20 of 81\%.
        \textit{Right:} Period predictions from noise-injected data, where recovery is significantly worse. Predicted periods have mean absolute percentage error of 246\%, median absolute percentage error of 24\%, acc10 of 28\%, and acc20 of 45\%. The horizontal band at 90 days represents targets where the model struggled to predict the period. In these cases, the prediction was biased toward the distribution median, or 90 days.}
        \label{fig:noise_comparison}
\end{figure*}

Our neural network's predictions on noiseless test data are shown in the left panel of Figure~\ref{fig:noise_comparison}.
The predicted periods have a mean absolute percentage error of 14\% and a median absolute percentage error of 7\%.
61\% of periods were successfully recovered to within 10\%, setting the bar for comparison to results for the noise-injected data.

\subsection{CNN performance on noisy data}
We present the neural network predictions on the noise-injected test data in the right panel of Figure~\ref{fig:noise_comparison}.
The predicted periods have a mean absolute percentage error of 246\% and a median absolute percentage error of 24\%.
Only 28\% of periods are successfully recovered to within 10\%.
The horizontal band at predicted period of 90 days represents simulated stars for which the network could not predict the period at all, instead assigning it the median of the period range.

\begin{figure*}
    \centering
    \includegraphics[width=\linewidth]{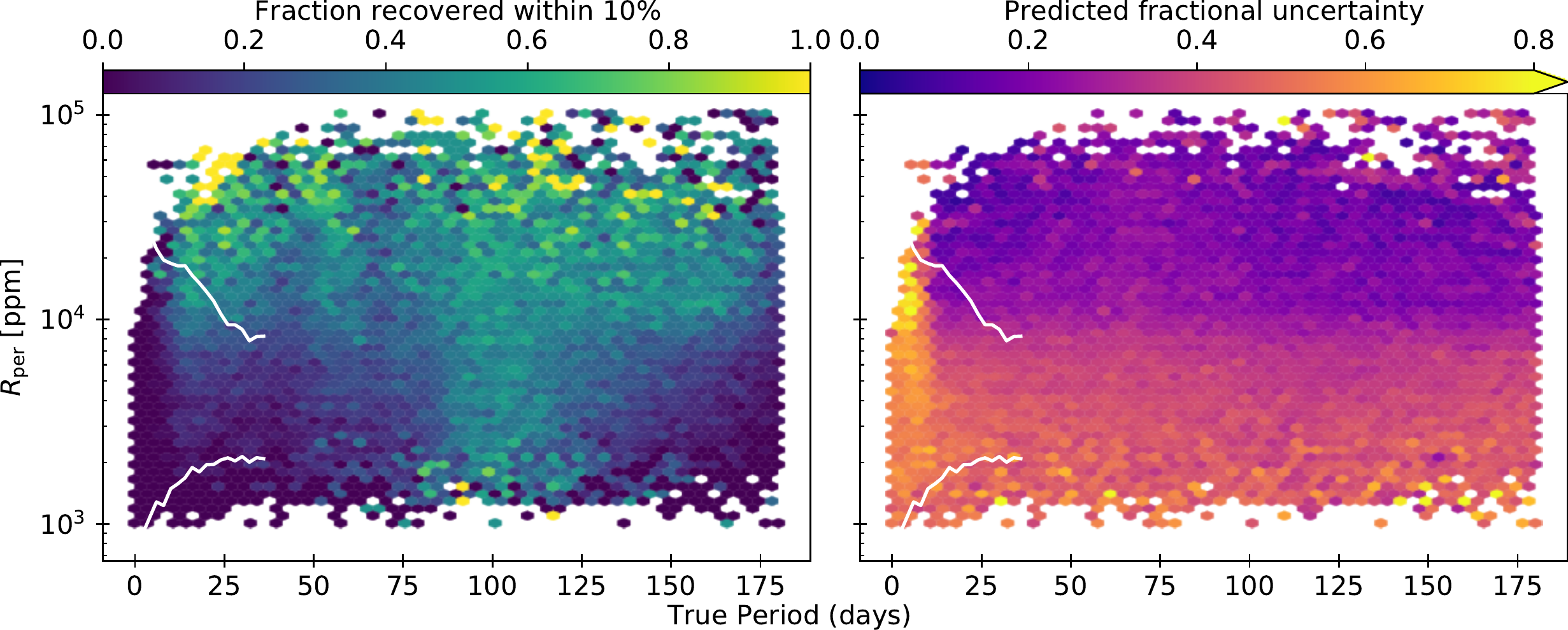}
    \caption{
        Neural network performance across the full simulation space of periods and amplitudes. In both panels, the white lines represent the 10th and 90th percentiles of the distributions from \citet{McQuillan2014}, to gauge where stars from \textit{Kepler} would fall.
        \textit{Left:} Period recovery rate as a function of period and amplitude for the noise-injected data. The neural network performs better at higher amplitudes, where the rotation signal overpowers instrumental noise.
        \textit{Right:} The same data, now colored by the neural-network-predicted fractional uncertainty in rotation period. The prediction is more certain for higher amplitudes. Furthermore, the prediction is most certain in the region with the highest recovery rate, indicating the predicted uncertainty is a reliable metric for period recovery without already knowing the true period.}
    \label{fig:amplitudes}
\end{figure*}

The addition of \textit{TESS}-like noise severely inhibits the performance of the neural network.
Like the conventional methods, the neural network predictions are more accurate at shorter periods.
When limiting to periods of 50 days or less, the median absolute percentage error is 12\%, and 44\% of targets are recovered to within 10\%.
The introduction of noise to the light curves also affects the amplitudes at which the network is most reliably predictive.
The left panel of Figure~\ref{fig:amplitudes} shows network recovery rate as a function of amplitude $R_\mathrm{per}$ \citep[as defined by][]{Basri2011} and equatorial rotation period.
Here, ``recovered" means the prediction is within 10\% of the true period.
As expected, the network performs better with higher-amplitude modulations, where the stellar signals are more easily picked out of the noise.

In addition to the rotation period, our choice of loss function allows us to predict the period uncertainty.
This value is a metric for how well the network is predicting the period.
The right panel of Figure~\ref{fig:amplitudes} shows the predicted uncertainty versus period and amplitude.
The predicted uncertainty, like the recovery rate, is better at higher amplitudes.
Since the predicted uncertainty correlates with the recovery rate, the predicted uncertainty is a reliable metric for successful period recovery without already knowing the period.
We can then use the predicted uncertainty to select a part of the sample recovered to a desired accuracy.

\begin{deluxetable*}{l|cccc|cccc}
\tablecaption{Metrics of Period Recovery on Simulated Light Curves}
\tablehead{\colhead{} &\multicolumn{4}{c}{$P_\mathrm{max} = 180$ days}& \multicolumn{4}{c}{$P_\mathrm{max} = 50$ days} \\
\hline
\colhead{Method} & \colhead{MAPE} & \colhead{MedAPE} & \colhead{acc10} & \colhead{acc20} & \colhead{MAPE} & \colhead{MedAPE} & \colhead{acc10} & \colhead{acc20} \\ 
\colhead{} & \colhead{(\%)} & \colhead{(\%)} & \colhead{(\%)} & \colhead{(\%)} & \colhead{(\%)} & \colhead{(\%)} & \colhead{(\%)} & \colhead{(\%)}} 
\startdata
LSP (noiseless) &       169 & 10 & 50 & 69 & 151 &  7 & 64 & 83 \\
GWPS (noiseless) &       51 &  8 & 56 & 77 &  51 &  6 & 67 & 85 \\
ACF (noiseless) &        31 &  6 & 63 & 82 &  50 &  5 & 69 & 87 \\
CNN (noiseless) &        14 &  7 & 61 & 81 &  11 &  5 & 69 & 86 \\
\hline
LSP (noisy) &            93 & 63 & 18 & 25 & 166 & 13 & 43 & 59 \\
GWPS (noisy) &           69 & 73 & 15 & 21 &  60 & 14 & 43 & 56 \\
ACF (noisy) &            94 & 53 & 26 & 37 & 190 & 25 & 36 & 47 \\
CNN (noisy, uncut) &    246 & 24 & 28 & 45 & 149 & 12 & 44 & 64 \\
CNN (noisy, cut) &       57 & 11 & 46 & 69 &  11 &  7 & 63 & 86 \\
\enddata
\label{tab:results}
\tablecomments{Recovery metrics for both the full 0.1--180-day period set, and for the subset with $P_\mathrm{rot} \leq 50$ days. All methods perform better on shorter-period stars, but our neural network consistently outperforms the conventional techniques on simulated light curves with real \textit{TESS} systematics.}
\end{deluxetable*}

\begin{figure*}
    \centering
    \includegraphics[width=0.7\linewidth]{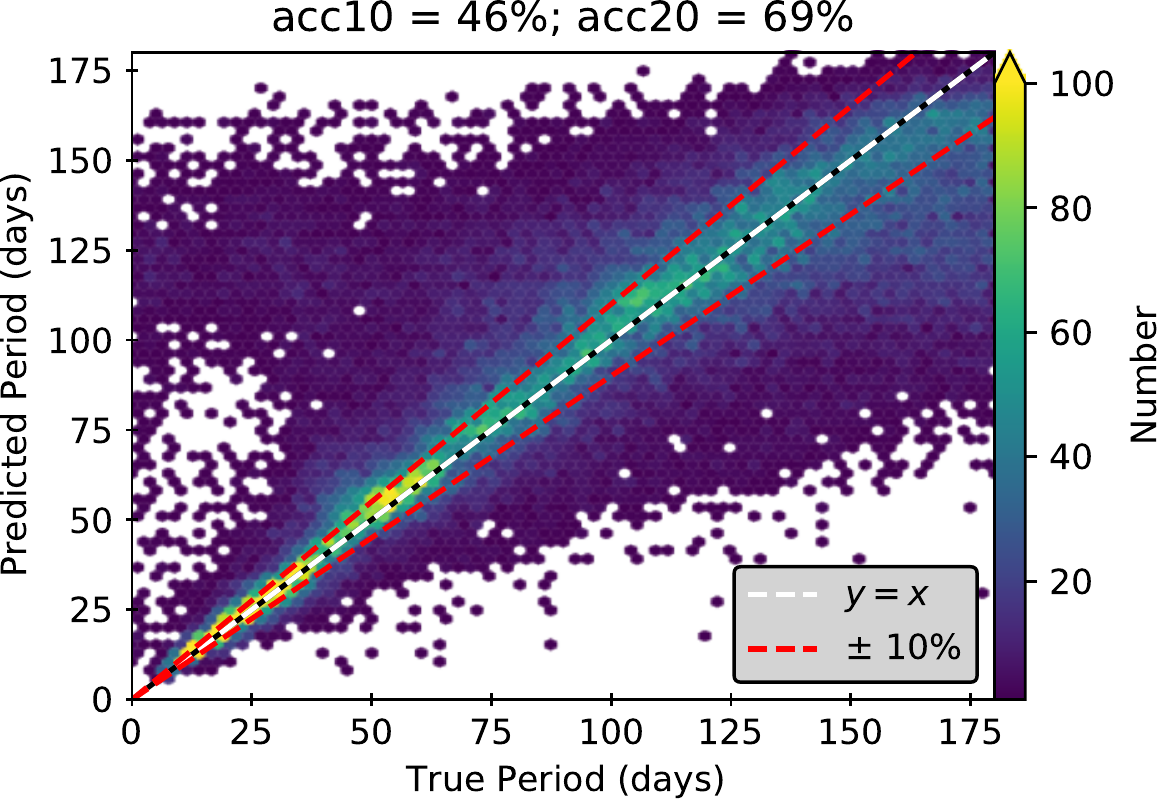}
    \caption{Period recovery for the half of the test set with the lowest predicted fractional uncertainty. Predicted periods have mean absolute percentage error of 57\%, median absolute percentage error of 11\%, acc10 of 46\%, and acc20 of 69\%. The predicted error cut removes the cluster of predicted periods at 90 days, giving credence to our cut to remove spurious period predictions. The cloud of objects with short true periods and predicted periods between 100 and 150 days have low fractional error because their predicted periods are large, but they account for only 4\% of the objects remaining after the cut.}
    \label{fig:filtered_prediction}
\end{figure*}

For our analysis, we selected the half of the test set with the lowest predicted fractional uncertainty.
The median predicted uncertainty for the sample before the cut was $\sigma_\mathrm{pred}/P_\mathrm{pred} = 0.35$.
The period recovery for the best-predicted half of the sample is shown in Figure~\ref{fig:filtered_prediction}.
The cut removed the horizontal band at predicted period of 90 days, and all summary statistics were improved.
46\% of periods were correctly predicted to within 10\%, and 69\% were accurate to within 20\%.
The predicted periods had a mean absolute percentage error of 57\% and a median absolute percentage error of 11\%.
A few targets with incorrectly predicted period between 100 and 150 days remained after the cut.
These had low predicted fractional uncertainty due to their large predicted period compared, so they made the cut despite being poorly predicted.
They accounted for about 4\% of the sample after the cut.

As with all other methods, the CNN performed better on the noisy data when limited to targets with periods less than 50 days.
For this subset of the sample, the median predicted fractional uncertainty was $\sigma_\mathrm{pred}/P_\mathrm{pred} = 0.2$.
Making the same cut as before (using the median fractional uncertainty of 0.2), the recovery of short-period stars improved to acc10 of 58\% and median absolute percentage error of 8\%.
Table~\ref{tab:results} shows the complete summary of our recovery results.

\section{Discussion}
We have demonstrated that convolutional neural networks are capable of extracting period information from noisy light curves or, more precisely, transformations of noisy light curves.
Our model also predicts the uncertainty in the period estimate, enabling us to see where the network is most successful and determine which period predictions are most reliable.
Here we discuss the strengths and weaknesses of our approach and compare them to those of conventional period detection methods.
We then comment on the prospects of estimating rotation periods from real \textit{TESS} light curves.

\subsection{Strengths and weaknesses of deep learning approach}
Our CNN outperformed conventional techniques in the recovery of rotation periods for the same underlying sample of synthetic light curves.
Whereas the conventional methods failed to recover periods longer than $\sim$2 \textit{TESS} sectors, our method successfully recovered simulated star periods across the full simulation range of 0.1--180 days.
Simulated stars in the range of periods yet unprobed by \textit{TESS}---13.7 days up to 90 days and beyond---were recovered with the highest success rate.
The recovery rate trails off at each end of the range ($P_\mathrm{rot} < 10$~days and $P_\mathrm{rot} > 170$~days) because of the choice of loss function: predicting the median under uncertainty biases predictions toward the median of the ensemble distribution and away from the ends of the range.

The challenge for classic period-recovery methods in \textit{TESS} light curves is mostly due to sector-to-sector stitching and the presence of scattered moonlight (repeating every 27 and 13.7 days, respectively).
Other effects such as temperature changes and momentum dumps appear at periods of 1.5, 2, 2.5, 3, 5, and 13.7 days \citep{TESSHandbook}.
All these effects combine to leave periodic imprints in the data that dominate stellar rotation signals and are difficult to remove.
All three of our conventional method tests significantly misidentified 27-days as the rotation period. Different methods latch onto different signals as well.
For example, ACF has significant misidentifications at 2.5 and 13.7 days and a weak twice-period alias, while WPS mistakes 1- and 5-day signals as the rotation period.
LSP mistakes these high-frequency signals less often, but often falls prey to half-period aliases, as does WPS.

It is noteworthy that, unlike with LSP and WPS, our neural network has no significant misidentification of half-period aliases or the high-frequency systematic aliases.
This is especially encouraging since we use WPS as the basis for our training data.
The fact that these aliases certainly exist in the training set but are not chosen as the period supports our claim that neural networks can learn and bypass systematic and false-period signals.
At the very least, if the rotation period is ambiguous, the network will predict a large uncertainty, allowing us to throw away the prediction.
Our results suggest that convolutional neural networks can learn systematic effects and regress rotation periods despite them, a significant step towards enabling large stellar rotation studies with \textit{TESS}.

\subsection{Comparisons to other period recovery attempts}
Our results suggest that this method would be on par with or better than other recent attempts to estimate $>13$-day rotation periods from \textit{TESS} light curves. 
\citet{CantoMartins2020} used a combination of Fast Fourier Transform, Lomb-Scargle, and wavelet techniques to estimate periods for 1,000 \textit{TESS} objects of interest. 
They obtained unambiguous rotation periods for 131 stars, but all were shorter than the 13.7-days \textit{TESS} orbital period.

\citet{Lu2020} trained a random forest (RF) regressor to predict rotation periods from 27-day sections of \textit{Kepler} light curves coupled with \textit{Gaia} stellar parameters. 
They then evaluated the trained model on single sectors of \textit{TESS} data for the same stars. 
Despite the stark differences in light curve systematics, they were able to recover rotation periods up to $\sim$50 days with 55\% accuracy, which is on par with the 57\% mean uncertainty achieved by our model. 
There are caveats to this comparison, however. First, the RF regressor relied primarily on effective temperature and secondarily on the light curve variability amplitude; light curve periodicity was not used for the period regression. 
Second, \citet{Lu2020} used two-minute cadence, Pre-search Data Conditioned Simple Aperture Photometry (PDCSAP) \textit{TESS} light curves, while our light curves were thirty-minute cadence, and our processing pipeline was more similar to simple aperture photometry (SAP). 
PDCSAP light curves are subjected to much heavier detrending than those produced by SAP methods. 
Finally, \citet{Lu2020} used real \textit{TESS} data, while we used simulated light curves. 
Each set comprises different distributions of rotation period, amplitude, and other important parameters. 
Because of these caveats, we encourage the reader to take caution when comparing the results of these studies.

\subsection{Prospects for measuring periodicity in \textit{TESS}}

We have so far demonstrated the ability to recover photometric rotation periods from simulated \textit{TESS}-like stellar light curves using deep learning. 
But the biggest question remains: can we reliably measure long periods from real \textit{TESS} data?

This is a difficult question to answer definitively for several reasons.
First, validation of any method requires a set of real stars for which rotation periods are already known.
The ideal data set for comparison is \textit{Kepler}, where tens of thousands of periods have been recorded \citep{McQuillan2014, Santos2019}.
Unfortunately, the overlap between \textit{TESS} and \textit{Kepler} is small: most \textit{Kepler} stars were observed for only a single sector at a time in \textit{TESS}.
With only a 27-day baseline, it is impossible to validate a method of obtaining long periods.
Stars in the \textit{TESS} CVZs were monitored continuously for almost a year, but only a handful of these stars have previously known rotation periods.

Despite the limitations, we attempted to recover rotation periods for a handful of stars observed by \textit{Kepler}, the Kilodegree Extremely Little Telescope survey \citep[KELT,][]{Pepper2007}, the MEarth Project \citep{Berta2012}, and the All-Sky Automated Survey for Supernovae \citep[ASAS-SN][]{Shappee2014, Kochanek2017}.

\subsubsection{\textit{Kepler}}
We targeted the few stars in the \textit{Kepler} field that had two consecutive sectors in \textit{TESS}, offering a baseline of roughly 50 days.
We simulated an entirely new training set with periods spanning 0.1 to 50 days, using a sample of galaxies in the \textit{Kepler} field as the noise model. 
With a 50-day baseline, only periods of up to 25 days might be recoverable, as timescales longer than this may be dominated by edge effects in the wavelet transform.
Even so, our network was unable to recover \textit{Kepler} periods reliably.

\subsubsection{KELT}
\begin{figure*}
    \centering
    \includegraphics{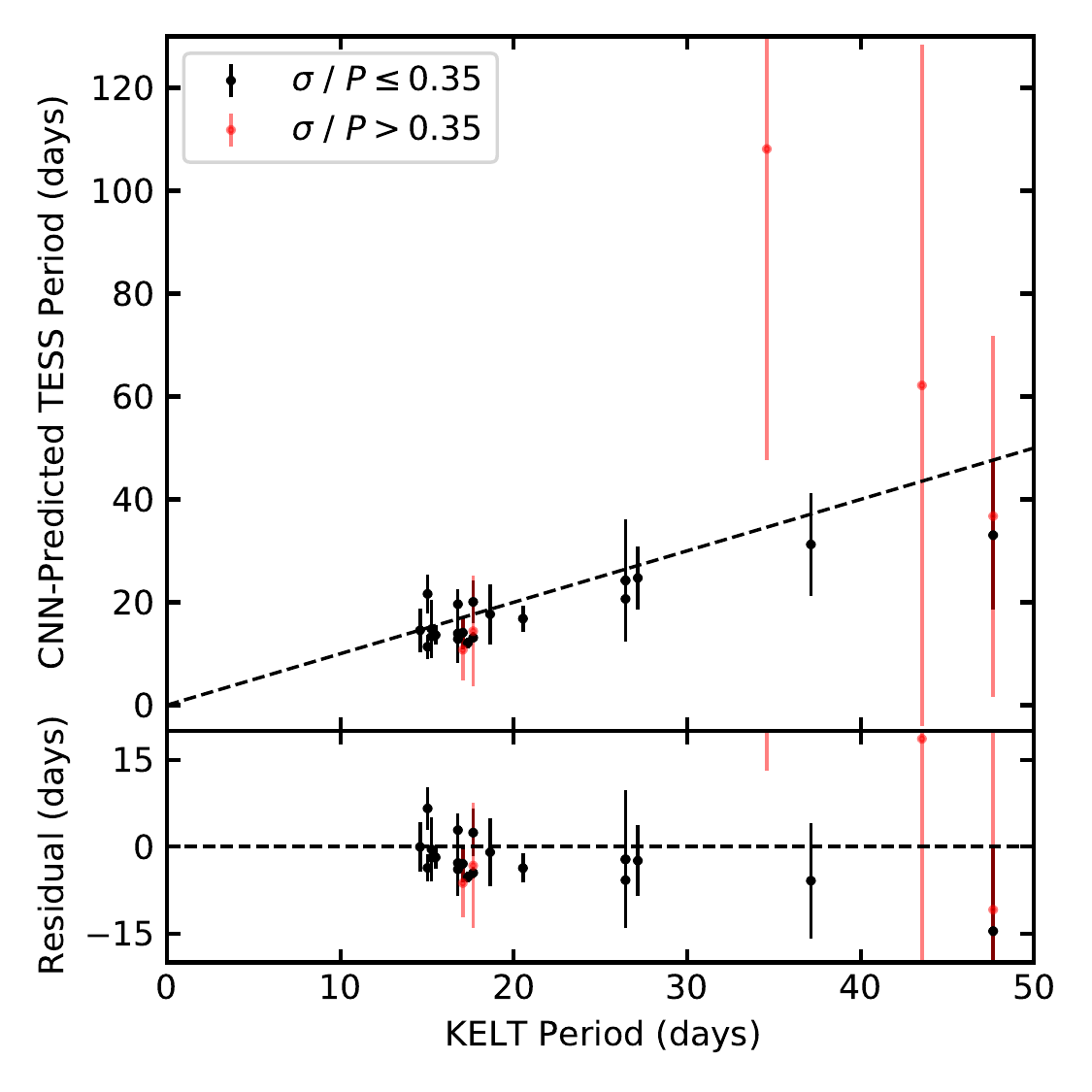}
    \caption{Period recovery of stars in both \textit{TESS} and KELT for which rotational modulation is apparent in the light curve. We applied the same fractional uncertainty cut applied to the simulation recovery results; the 21 stars that made the cut are in black, while the 5 stars with unreliable period predictions are in red. We successfully recovered periods longer than 13.7 and even 27 days from real \textit{TESS} light curves. Even when \textit{TESS} aliases are the dominant sources of power in the wavelet transform, our neural network was able to recover the correct rotation period.}
    \label{fig:kelt_recovery}
\end{figure*}

We similarly targeted 106 stars in the \textit{TESS} SCVZ also observed by the KELT survey. 
\citet{Oelkers2018b} obtained rotation periods for these stars using Lomb-Scargle periodograms of their KELT light curves.
We specifically selected stars with a measured period greater than 13.7 days to test recovery of long periods.
To maximize the chances of recovering rotation periods, we used the \textit{TESS} Science Processing Operations Center (SPOC) FFI simple aperture photometry (SAP) light curves \citep{Caldwell2020}.
At the time of writing, only sectors 1-6 were available, but we trained our CNN using year-long (13-sector) light curves.
However, the construction of our wavelet power spectrum used the same vertical (frequency) axis regardless of light curve length, so any length of light curve could be used without needing to retrain the neural network.

Upon visual inspection of the \textit{TESS}-SPOC light curves, we noticed that many did not show obvious rotational modulation.
Selecting only those light curves with unambiguous rotational modulation, we were left with 26 light curves with KELT rotation periods spanning 13.7 to 47 days.
We generated wavelet power spectra and evaluated our neural network on them.
Figure~\ref{fig:kelt_recovery} shows our predictions for these 26 KELT stars. 
We subjected these points to the same cut in predicted fractional uncertainty as in Figure~\ref{fig:filtered_prediction}. 
Stars that made the cut are displayed in black, while stars whose predicted uncertainties were too large are shown in red. 
We successfully recovered stars with rotation periods longer than 13.7 and 27 days, even when \textit{TESS} systematics were the dominant source of power in the wavelet power spectra.
Furthermore, using the predicted fractional uncertainty as a quality cut successfully removed stars whose predictions were unreliable or wrong while ensuring the most reliable predictions remained in the sample.

\subsubsection{MEarth and TOI-700}
Only one long-period target was observed by MEarth in the \textit{TESS} SCVZ: TIC 149423103. 
\citet{Newton2018} measured a rotation period of 111 days for this target from MEarth data. 
Using our neural network on the FFI data from \textit{TESS}, we obtained $P_\mathrm{rot} = 116 \pm 48$~days. 
While the predicted period was within 5\% of the ``true" period, the relatively large uncertainty (41\%) means this target would fail our quality cut, and an ensemble period recovery attempt would miss it.

TOI-700 is another well-characterized star in the SCVZ. 
Using ASAS-SN data, \citet{Gilbert2020} estimated a precise rotation period of 54.0$\pm$0.8 days. 
\citet{Hedges2020} used a systematics-insensitive periodogram of its \textit{TESS} light curve to obtain a period of 52.8 days. 
With our model we predicted a period of 59$\pm$53 days. 
Our period prediction was accurate to within 10\%, but the large uncertainty would cause this target to be missed as well.

\subsubsection{General period recovery and improvements}
While we successfully recovered the rotation periods of these few hand-picked stars, robust recovery on larger, statistical scales will require more work and vetting. 
Our method allows us to see beyond the 13.7-day barrier, but only the stars with the largest amplitudes were reliably recovered---we are still limited by the \textit{TESS} noise floor.
\textit{TESS} is less precise than \textit{Kepler} at all magnitudes \citep{TESSHandbook}, so spot modulations require higher amplitudes to rise above the noise.
In our simulated test set, we recovered rotation signals with some success down to amplitudes of a few parts per thousand, but our model was most successful at and above amplitudes of 1\%. 
Both panels of Figure~\ref{fig:amplitudes} show the 10th- and 90th- percentile envelopes of the rotating stars from \citet{McQuillan2014}. 
The bulk of \textit{Kepler}'s rotating population falls between amplitudes of 1 and 10 parts per thousand and lie in a region where our recovery was less successful. 
These kinds of stars will be more difficult to recover with \textit{TESS}, whatever the method.

Still, we believe improvements to our method will maximize what is recoverable from \textit{TESS}. 
There are several ways to extend the predictability of our neural network to lower amplitudes and enhance the predictability at high amplitudes.
The first and perhaps most useful improvement will come through light curve processing. 
Our processing pipeline followed the regression corrector documentation of \citet{LightkurveDocs} using a magnitude-dependent aperture threshold. 
In practice, a more carefully developed pipeline should be preferred. 
At the time of writing, the FFIs of sectors 1-6 have been reduced by both the \textit{TESS} Asteroseismic Science Operations Center (TASOC) pipeline and the \textit{TESS} Science Processing Operations Center (SPOC) pipeline. 
Once sectors 7-13 are processed, the Southern Continuous Viewing Zone (SCVZ) will be complete, providing year-long light curves for hundreds of thousands of targets.
These light curves will feature more careful systematics removal and should contain cleaner examples to use as "pure noise" light curves in our sample.
We leave the use of these light curves to a future paper.

Another improvement may come with the inclusion of observation metadata. 
At least with \textit{TESS}, certain systematic effects are specific to particular cameras, CCDs, or even locations on a CCD.
Including camera number, CCD number, and x- and y- pixel coordinates in the training data set will allow neural networks to learn where to expect certain features and more easily ignore them in favor of astrophysical signals.

Improvements can be made to the neural network as well. In its current form, our model assumes that all input signals have rotation signatures, but not all real light curves display rotational modulation. In the future we may include a classification step like \citet{Lu2020} to determine which signals contain rotational modulation. Adding this classification step will allow the regressor to focus on signals with recoverable rotation, making for more efficient training. 

It is important to note that our implementations of the conventional period recovery techniques still perform better than in reality \citep[e.g.,][who were unsuccessful in recovering anything past 13.7 days]{CantoMartins2020, Avallone2021}.
This indicates that, despite all our attempts to create as realistic a training set as possible, our simulations are not perfectly representative of real stars.
It could be that our stitching routine fails to suppress long-period signals as the real light curves do.
Another possibility is that our spot model, while tuned to the Sun, may not be representative of real spots on other stars.
Whatever the reason, we have demonstrated the ability to recover periods even when the systematics that \textit{are} present in our simulations make conventional techniques fail.

Even though our spot evolution simulations include latitudinal differential rotation, we were unable to recover differential rotation in this study.
In some wavelet power spectra of our simulated light curves, the differential rotation is apparent as a slope in the frequency of maximum power versus time.
When binning the power spectra to $64\times64$ pixels, the slope is more difficult to resolve.
While increasing the resolution of the wavelet power spectrum images should enable the recovery of differential rotation, it will come at the expense of longer training time.
We will investigate the recovery of differential rotation, activity levels, and spot properties in future work.

If we can see beyond the complicated systematics, \textit{TESS} will deliver the largest set of rotation periods yet.
\citet{McQuillan2014} obtained rotation periods for 34,000 stars in the \textit{Kepler} field.
The \textit{TESS} continuous viewing zones combine to cover 900 square degrees around the ecliptic poles, representing about eight times the sky coverage of \textit{Kepler} during its primary mission.
We can therefore expect hundreds of thousands of new stars with rotation period estimates from the \textit{TESS} CVZs, and perhaps more from lower ecliptic latitudes.
Because of \textit{TESS}'s lower precision compared to \textit{Kepler}, the true number will likely be somewhat smaller, but the prospect of hundreds of thousands of new periods is worth continued refinements of this technique.
We leave the application of this tool to the full CVZ samples to a future paper.

\section{Summary and Conclusion}

We used a convolutional neural network to recover rotation periods and uncertainties from simulated light curves with real \textit{TESS} systematics.
Despite the systematics, we successfully recovered periods even for targets whose periods were longer than the 13.7-day barrier encountered by conventional period recovery methods.
In the half of the simulated test data with the smallest predicted fractional uncertainty, we recovered 10\%-accurate periods for 46\% of the sample, and 20\%-accurate periods for 69\% of the sample.
We also found no significant misidentification of half-period aliases, unlike the Lomb-Scargle and wavelet methods.
While periods were retrieved more successfully from higher-amplitude signals, the ability to predict uncertainties allows us to probe lower-amplitude rotation signals as well.

In future work, we plan to use this method to produce a catalog of rotation periods from \textit{TESS} full-frame image light curves.
We will also add output options to our neural network to predict latitudinal differential rotation and understand more of the properties used to produce the training set.
With deep learning, we hope to maximize the output of \textit{TESS} in spite of the frustrations that arise from its systematics.
The ability to recover rotation periods, especially long periods, from \textit{TESS} data will finally enable large studies of rotation across diverse populations of stars in the Galaxy if only the systematics can be learned.

\begin{acknowledgments}

The authors wish to acknowledge Gagandeep Anand, Ashley Chontos, Aidan Chun, Curt Dodds, Ryan Dungee, Kyle Hart, Rae Holcomb, Daniel Huber, Miles Lucas, Sushant Mahajan, Anna Payne, Nicholas Saunders, Benjamin Shappee, Xudong Sun, and Jamie Tayar for fruitful conversations that improved the quality of this work.

The technical support and advanced computing resources from the University of Hawai‘i Information Technology Services – Cyberinfrastructure are gratefully acknowledged.

This research was supported in part by the National Science Foundation under Grant No. NSF PHY-1748958.

J.v.S. and Z.R.C. acknowledge support from the National Aeronautics and Space Administration (80NSSC21K0246, 80NSSC18K18584)

J.L. acknowledges support from NASA through an Astrophysics Data Analysis Program grant to Lowell Observatory (grant 80NSSC20K1001).

This paper includes data collected by the \textit{TESS} mission. Funding for the \textit{TESS} mission is provided by the NASA's Science Mission Directorate.

\end{acknowledgments}

\software{\texttt{NumPy} \citep{Numpy2020}, \texttt{Pandas} \citep{Pandas2010}, \texttt{Matplotlib} \citep{Matplotlib2007}, \texttt{AstroPy} \citep{Astropy2013, Astropy2018}, \texttt{SciPy} \citep{Scipy2020}, \texttt{PyTorch} \citep{Pytorch2019}, \texttt{Lightkurve} \citep{Lightkurve2018}, \texttt{TESScut} \citep{Brasseur2019}, \texttt{iPython} \citep{iPython2007}, \texttt{butterpy} \citep{butterpy}, \texttt{starspot} \citep{Starspot2021}}

\bibliography{references}

\begin{thebibliography}{}
\expandafter\ifx\csname natexlab\endcsname\relax\def\natexlab#1{#1}\fi
\providecommand{\url}[1]{\href{#1}{#1}}
\providecommand{\dodoi}[1]{doi:~\href{http://doi.org/#1}{\nolinkurl{#1}}}
\providecommand{\doeprint}[1]{\href{http://ascl.net/#1}{\nolinkurl{http://ascl.net/#1}}}
\providecommand{\doarXiv}[1]{\href{https://arxiv.org/abs/#1}{\nolinkurl{https://arxiv.org/abs/#1}}}

\bibitem[{{Aigrain} {et~al.}(2012){Aigrain}, {Pont}, \& {Zucker}}]{Aigrain2012}
{Aigrain}, S., {Pont}, F., \& {Zucker}, S. 2012, \mnras, 419, 3147,
  \dodoi{10.1111/j.1365-2966.2011.19960.x}

\bibitem[{{Aigrain} {et~al.}(2015){Aigrain}, {Llama}, {Ceillier}, {Chagas},
  {Davenport}, {Garc{\'\i}a}, {Hay}, {Lanza}, {McQuillan}, {Mazeh}, {de
  Medeiros}, {Nielsen}, \& {Reinhold}}]{Aigrain2015}
{Aigrain}, S., {Llama}, J., {Ceillier}, T., {et~al.} 2015, \mnras, 450, 3211,
  \dodoi{10.1093/mnras/stv853}

\bibitem[{{Amard} {et~al.}(2020){Amard}, {Roquette}, \& {Matt}}]{Amard2020}
{Amard}, L., {Roquette}, J., \& {Matt}, S.~P. 2020, \mnras, 499, 3481,
  \dodoi{10.1093/mnras/staa3038}

\bibitem[{{Angus}(2021)}]{Starspot2021}
{Angus}, R. 2021, {starspot: code for measuring stellar rotation periods},
  v0.2,  Zenodo, \dodoi{10.5281/zenodo.4613887}.
\newblock \url{https://doi.org/10.5281/zenodo.4613887}

\bibitem[{{Angus} {et~al.}(2018){Angus}, {Morton}, {Aigrain}, {Foreman-Mackey},
  \& {Rajpaul}}]{Angus2018}
{Angus}, R., {Morton}, T., {Aigrain}, S., {Foreman-Mackey}, D., \& {Rajpaul},
  V. 2018, \mnras, 474, 2094, \dodoi{10.1093/mnras/stx2109}

\bibitem[{{Astropy Collaboration} {et~al.}(2013){Astropy Collaboration},
  {Robitaille}, {Tollerud}, {Greenfield}, {Droettboom}, {Bray}, {Aldcroft},
  {Davis}, {Ginsburg}, {Price-Whelan}, {Kerzendorf}, {Conley}, {Crighton},
  {Barbary}, {Muna}, {Ferguson}, {Grollier}, {Parikh}, {Nair}, {Unther},
  {Deil}, {Woillez}, {Conseil}, {Kramer}, {Turner}, {Singer}, {Fox}, {Weaver},
  {Zabalza}, {Edwards}, {Azalee Bostroem}, {Burke}, {Casey}, {Crawford},
  {Dencheva}, {Ely}, {Jenness}, {Labrie}, {Lim}, {Pierfederici}, {Pontzen},
  {Ptak}, {Refsdal}, {Servillat}, \& {Streicher}}]{Astropy2013}
{Astropy Collaboration}, {Robitaille}, T.~P., {Tollerud}, E.~J., {et~al.} 2013,
  \aap, 558, A33, \dodoi{10.1051/0004-6361/201322068}

\bibitem[{{Astropy Collaboration} {et~al.}(2018){Astropy Collaboration},
  {Price-Whelan}, {Sip{\H{o}}cz}, {G{\"u}nther}, {Lim}, {Crawford}, {Conseil},
  {Shupe}, {Craig}, {Dencheva}, {Ginsburg}, {Vand erPlas}, {Bradley},
  {P{\'e}rez-Su{\'a}rez}, {de Val-Borro}, {Aldcroft}, {Cruz}, {Robitaille},
  {Tollerud}, {Ardelean}, {Babej}, {Bach}, {Bachetti}, {Bakanov}, {Bamford},
  {Barentsen}, {Barmby}, {Baumbach}, {Berry}, {Biscani}, {Boquien}, {Bostroem},
  {Bouma}, {Brammer}, {Bray}, {Breytenbach}, {Buddelmeijer}, {Burke},
  {Calderone}, {Cano Rodr{\'\i}guez}, {Cara}, {Cardoso}, {Cheedella}, {Copin},
  {Corrales}, {Crichton}, {D'Avella}, {Deil}, {Depagne}, {Dietrich}, {Donath},
  {Droettboom}, {Earl}, {Erben}, {Fabbro}, {Ferreira}, {Finethy}, {Fox},
  {Garrison}, {Gibbons}, {Goldstein}, {Gommers}, {Greco}, {Greenfield},
  {Groener}, {Grollier}, {Hagen}, {Hirst}, {Homeier}, {Horton}, {Hosseinzadeh},
  {Hu}, {Hunkeler}, {Ivezi{\'c}}, {Jain}, {Jenness}, {Kanarek}, {Kendrew},
  {Kern}, {Kerzendorf}, {Khvalko}, {King}, {Kirkby}, {Kulkarni}, {Kumar},
  {Lee}, {Lenz}, {Littlefair}, {Ma}, {Macleod}, {Mastropietro}, {McCully},
  {Montagnac}, {Morris}, {Mueller}, {Mumford}, {Muna}, {Murphy}, {Nelson},
  {Nguyen}, {Ninan}, {N{\"o}the}, {Ogaz}, {Oh}, {Parejko}, {Parley}, {Pascual},
  {Patil}, {Patil}, {Plunkett}, {Prochaska}, {Rastogi}, {Reddy Janga},
  {Sabater}, {Sakurikar}, {Seifert}, {Sherbert}, {Sherwood-Taylor}, {Shih},
  {Sick}, {Silbiger}, {Singanamalla}, {Singer}, {Sladen}, {Sooley},
  {Sornarajah}, {Streicher}, {Teuben}, {Thomas}, {Tremblay}, {Turner},
  {Terr{\'o}n}, {van Kerkwijk}, {de la Vega}, {Watkins}, {Weaver}, {Whitmore},
  {Woillez}, {Zabalza}, \& {Astropy Contributors}}]{Astropy2018}
{Astropy Collaboration}, {Price-Whelan}, A.~M., {Sip{\H{o}}cz}, B.~M., {et~al.}
  2018, \aj, 156, 123, \dodoi{10.3847/1538-3881/aabc4f}

\bibitem[{{Avallone} {et~al.}(2021, in prep.){Avallone}, {Tayar}, {van Saders},
  {Berger}, \& {Claytor}}]{Avallone2021}
{Avallone}, E.~A., {Tayar}, J., {van Saders}, J.~L., {Berger}, T.~A., \&
  {Claytor}, Z.~R. 2021, in prep.

\bibitem[{{Baglin} {et~al.}(2006){Baglin}, {Auvergne}, {Barge}, {Deleuil},
  {Catala}, {Michel}, {Weiss}, \& {COROT Team}}]{Baglin2006}
{Baglin}, A., {Auvergne}, M., {Barge}, P., {et~al.} 2006, in ESA Special
  Publication, Vol. 1306, The CoRoT Mission Pre-Launch Status - Stellar
  Seismology and Planet Finding, ed. M.~{Fridlund}, A.~{Baglin}, J.~{Lochard},
  \& L.~{Conroy}, 33

\bibitem[{{Basri} {et~al.}(2011){Basri}, {Walkowicz}, {Batalha}, {Gilliland},
  {Jenkins}, {Borucki}, {Koch}, {Caldwell}, {Dupree}, {Latham}, {Marcy},
  {Meibom}, \& {Brown}}]{Basri2011}
{Basri}, G., {Walkowicz}, L.~M., {Batalha}, N., {et~al.} 2011, \aj, 141, 20,
  \dodoi{10.1088/0004-6256/141/1/20}

\bibitem[{{Bazot} {et~al.}(2018){Bazot}, {Nielsen}, {Mary},
  {Christensen-Dalsgaard}, {Benomar}, {Petit}, {Gizon}, {Sreenivasan}, \&
  {White}}]{Bazot2018}
{Bazot}, M., {Nielsen}, M.~B., {Mary}, D., {et~al.} 2018, \aap, 619, L9,
  \dodoi{10.1051/0004-6361/201834251}

\bibitem[{{Berta} {et~al.}(2012){Berta}, {Irwin}, {Charbonneau}, {Burke}, \&
  {Falco}}]{Berta2012}
{Berta}, Z.~K., {Irwin}, J., {Charbonneau}, D., {Burke}, C.~J., \& {Falco},
  E.~E. 2012, \aj, 144, 145, \dodoi{10.1088/0004-6256/144/5/145}

\bibitem[{{Blancato} {et~al.}(2020){Blancato}, {Ness}, {Huber}, {Lu}, \&
  {Angus}}]{Blancato2020}
{Blancato}, K., {Ness}, M., {Huber}, D., {Lu}, Y., \& {Angus}, R. 2020, arXiv
  e-prints, arXiv:2005.09682.
\newblock \doarXiv{2005.09682}

\bibitem[{{Borucki} {et~al.}(2010){Borucki}, {Koch}, {Basri}, {Batalha},
  {Brown}, {Caldwell}, {Caldwell}, {Christensen-Dalsgaard}, {Cochran},
  {DeVore}, {Dunham}, {Dupree}, {Gautier}, {Geary}, {Gilliland}, {Gould},
  {Howell}, {Jenkins}, {Kondo}, {Latham}, {Marcy}, {Meibom}, {Kjeldsen},
  {Lissauer}, {Monet}, {Morrison}, {Sasselov}, {Tarter}, {Boss}, {Brownlee},
  {Owen}, {Buzasi}, {Charbonneau}, {Doyle}, {Fortney}, {Ford}, {Holman},
  {Seager}, {Steffen}, {Welsh}, {Rowe}, {Anderson}, {Buchhave}, {Ciardi},
  {Walkowicz}, {Sherry}, {Horch}, {Isaacson}, {Everett}, {Fischer}, {Torres},
  {Johnson}, {Endl}, {MacQueen}, {Bryson}, {Dotson}, {Haas}, {Kolodziejczak},
  {Van Cleve}, {Chandrasekaran}, {Twicken}, {Quintana}, {Clarke}, {Allen},
  {Li}, {Wu}, {Tenenbaum}, {Verner}, {Bruhweiler}, {Barnes}, \&
  {Prsa}}]{Borucki2010}
{Borucki}, W.~J., {Koch}, D., {Basri}, G., {et~al.} 2010, Science, 327, 977,
  \dodoi{10.1126/science.1185402}

\bibitem[{{Brasseur} {et~al.}(2019){Brasseur}, {Phillip}, {Fleming},
  {Mullally}, \& {White}}]{Brasseur2019}
{Brasseur}, C.~E., {Phillip}, C., {Fleming}, S.~W., {Mullally}, S.~E., \&
  {White}, R.~L. 2019, {Astrocut: Tools for creating cutouts of TESS images}.
\newblock \doeprint{1905.007}

\bibitem[{{Caldwell} {et~al.}(2020){Caldwell}, {Tenenbaum}, {Twicken},
  {Jenkins}, {Ting}, {Smith}, {Hedges}, {Fausnaugh}, {Rose}, \&
  {Burke}}]{Caldwell2020}
{Caldwell}, D.~A., {Tenenbaum}, P., {Twicken}, J.~D., {et~al.} 2020, Research
  Notes of the American Astronomical Society, 4, 201,
  \dodoi{10.3847/2515-5172/abc9b3}

\bibitem[{{Canto Martins} {et~al.}(2020){Canto Martins}, {Gomes}, {Messias},
  {de Lira}, {Le{\~a}o}, {Almeida}, {Teixeira}, {das Chagas}, {Bravo}, {Bewketu
  Belete}, \& {De Medeiros}}]{CantoMartins2020}
{Canto Martins}, B.~L., {Gomes}, R.~L., {Messias}, Y.~S., {et~al.} 2020, \apjs,
  250, 20, \dodoi{10.3847/1538-4365/aba73f}

\bibitem[{{Ceillier} {et~al.}(2017){Ceillier}, {Tayar}, {Mathur}, {Salabert},
  {Garc{\'\i}a}, {Stello}, {Pinsonneault}, {van Saders}, {Beck}, \&
  {Bloemen}}]{Ceillier2017}
{Ceillier}, T., {Tayar}, J., {Mathur}, S., {et~al.} 2017, \aap, 605, A111,
  \dodoi{10.1051/0004-6361/201629884}

\bibitem[{Claytor {et~al.}(2021)Claytor, Lucas, \& Llama}]{butterpy}
Claytor, Z.~R., Lucas, M., \& Llama, J. 2021, {Butterpy: realistic star spot
  evolution and light curves in Python}, 0.1.0,  Zenodo,
  \dodoi{10.5281/zenodo.4722052}.
\newblock \url{https://doi.org/10.5281/zenodo.4722052}

\bibitem[{{Claytor} {et~al.}(2020){Claytor}, {van Saders}, {Santos},
  {Garc{\'\i}a}, {Mathur}, {Tayar}, {Pinsonneault}, \&
  {Shetrone}}]{Claytor2020}
{Claytor}, Z.~R., {van Saders}, J.~L., {Santos}, {\^A}. R.~G., {et~al.} 2020,
  \apj, 888, 43, \dodoi{10.3847/1538-4357/ab5c24}

\bibitem[{Cranmer {et~al.}(2020)Cranmer, Brehmer, \& Louppe}]{Cranmer2020}
Cranmer, K., Brehmer, J., \& Louppe, G. 2020, Proceedings of the National
  Academy of Sciences, 117, 30055, \dodoi{10.1073/pnas.1912789117}

\bibitem[{{Davenport}(2017)}]{Davenport2017}
{Davenport}, J. R.~A. 2017, \apj, 835, 16, \dodoi{10.3847/1538-4357/835/1/16}

\bibitem[{{Feiden} {et~al.}(2011){Feiden}, {Guinan}, {Boyajian}, {Kok},
  {Basturk}, {Roberson}, \& {Ribas}}]{Feiden2011}
{Feiden}, G., {Guinan}, E., {Boyajian}, T., {et~al.} 2011, in American
  Astronomical Society Meeting Abstracts, Vol. 217, American Astronomical
  Society Meeting Abstracts \#217, 140.18

\bibitem[{{Feinstein} {et~al.}(2020){Feinstein}, {Montet}, {Ansdell}, {Nord},
  {Bean}, {G{\"u}nther}, {Gully-Santiago}, \& {Schlieder}}]{Feinstein2020}
{Feinstein}, A.~D., {Montet}, B.~T., {Ansdell}, M., {et~al.} 2020, arXiv
  e-prints, arXiv:2005.07710.
\newblock \doarXiv{2005.07710}

\bibitem[{{Flesch}(2015)}]{Flesch2015}
{Flesch}, E.~W. 2015, \pasa, 32, e010, \dodoi{10.1017/pasa.2015.10}

\bibitem[{{Garc{\'\i}a} {et~al.}(2014){Garc{\'\i}a}, {Ceillier}, {Salabert},
  {Mathur}, {van Saders}, {Pinsonneault}, {Ballot}, {Beck}, {Bloemen},
  {Campante}, {Davies}, {do Nascimento}, {Mathis}, {Metcalfe}, {Nielsen},
  {Su{\'a}rez}, {Chaplin}, {Jim{\'e}nez}, \& {Karoff}}]{Garcia2014}
{Garc{\'\i}a}, R.~A., {Ceillier}, T., {Salabert}, D., {et~al.} 2014, \aap, 572,
  A34, \dodoi{10.1051/0004-6361/201423888}

\bibitem[{{Gilbert} {et~al.}(2020){Gilbert}, {Barclay}, {Schlieder},
  {Quintana}, {Hord}, {Kostov}, {Lopez}, {Rowe}, {Hoffman}, {Walkowicz},
  {Silverstein}, {Rodriguez}, {Vanderburg}, {Suissa}, {Airapetian}, {Clement},
  {Raymond}, {Mann}, {Kruse}, {Lissauer}, {Col{\'o}n}, {Kopparapu},
  {Kreidberg}, {Zieba}, {Collins}, {Quinn}, {Howell}, {Ziegler}, {Vrijmoet},
  {Adams}, {Arney}, {Boyd}, {Brande}, {Burke}, {Cacciapuoti}, {Chance},
  {Christiansen}, {Covone}, {Daylan}, {Dineen}, {Dressing}, {Essack},
  {Fauchez}, {Galgano}, {Howe}, {Kaltenegger}, {Kane}, {Lam}, {Lee}, {Lewis},
  {Logsdon}, {Mandell}, {Monsue}, {Mullally}, {Mullally}, {Paudel},
  {Pidhorodetska}, {Plavchan}, {Reyes}, {Rinehart}, {Rojas-Ayala}, {Smith},
  {Stassun}, {Tenenbaum}, {Vega}, {Villanueva}, {Wolf}, {Youngblood}, {Ricker},
  {Vanderspek}, {Latham}, {Seager}, {Winn}, {Jenkins}, {Bakos}, {Brice{\~n}o},
  {Ciardi}, {Cloutier}, {Conti}, {Couperus}, {Di Sora}, {Eisner}, {Everett},
  {Gan}, {Hartman}, {Henry}, {Isopi}, {Jao}, {Jensen}, {Law}, {Mallia},
  {Matson}, {Shappee}, {Le Wood}, \& {Winters}}]{Gilbert2020}
{Gilbert}, E.~A., {Barclay}, T., {Schlieder}, J.~E., {et~al.} 2020, \aj, 160,
  116, \dodoi{10.3847/1538-3881/aba4b2}

\bibitem[{Goodfellow {et~al.}(2016)Goodfellow, Bengio, \&
  Courville}]{Goodfellow2016}
Goodfellow, I., Bengio, Y., \& Courville, A. 2016, Deep Learning (MIT Press)

\bibitem[{{Guiglion} {et~al.}(2020){Guiglion}, {Matijevic}, {Queiroz},
  {Valentini}, {Steinmetz}, {Chiappini}, {Grebel}, {McMillan}, {Kordopatis},
  {Kunder}, {Zwitter}, {Khalatyan}, {Anders}, {Enke}, {Minchev}, {Monari},
  {Wyse}, {Bienayme}, {Bland-Hawthorn}, {Gibson}, {Navarro}, {Parker}, {Reid},
  {Seabroke}, \& {Siebert}}]{Guiglion2020}
{Guiglion}, G., {Matijevic}, G., {Queiroz}, A.~B.~A., {et~al.} 2020, arXiv
  e-prints, arXiv:2004.12666.
\newblock \doarXiv{2004.12666}

\bibitem[{Harris {et~al.}(2020)Harris, Millman, van~der Walt, Gommers,
  Virtanen, Cournapeau, Wieser, Taylor, Berg, Smith, Kern, Picus, Hoyer, van
  Kerkwijk, Brett, Haldane, del R{'{\i}}o, Wiebe, Peterson,
  G{'{e}}rard-Marchant, Sheppard, Reddy, Weckesser, Abbasi, Gohlke, \&
  Oliphant}]{Numpy2020}
Harris, C.~R., Millman, K.~J., van~der Walt, S.~J., {et~al.} 2020, Nature, 585,
  357, \dodoi{10.1038/s41586-020-2649-2}

\bibitem[{{Hathaway}(2011)}]{Hathaway2011}
{Hathaway}, D.~H. 2011, \solphys, 273, 221, \dodoi{10.1007/s11207-011-9837-z}

\bibitem[{{Hathaway}(2015)}]{Hathaway2015}
---. 2015, Living Reviews in Solar Physics, 12, 4, \dodoi{10.1007/lrsp-2015-4}

\bibitem[{{Hathaway} {et~al.}(1994){Hathaway}, {Wilson}, \&
  {Reichmann}}]{Hathaway1994}
{Hathaway}, D.~H., {Wilson}, R.~M., \& {Reichmann}, E.~J. 1994, \solphys, 151,
  177, \dodoi{10.1007/BF00654090}

\bibitem[{{Hedges} {et~al.}(2020){Hedges}, {Angus}, {Barentsen}, {Saunders},
  {Montet}, \& {Gully-Santiago}}]{Hedges2020}
{Hedges}, C., {Angus}, R., {Barentsen}, G., {et~al.} 2020, Research Notes of
  the American Astronomical Society, 4, 220, \dodoi{10.3847/2515-5172/abd106}

\bibitem[{{Hezaveh} {et~al.}(2017){Hezaveh}, {Perreault Levasseur}, \&
  {Marshall}}]{Hezaveh2017}
{Hezaveh}, Y.~D., {Perreault Levasseur}, L., \& {Marshall}, P.~J. 2017, \nat,
  548, 555, \dodoi{10.1038/nature23463}

\bibitem[{{Holcomb}(2020)}]{Holcomb2020}
{Holcomb}, R. 2020, in American Astronomical Society Meeting Abstracts,
  American Astronomical Society Meeting Abstracts, 274.04

\bibitem[{{Hunter}(2007)}]{Matplotlib2007}
{Hunter}, J.~D. 2007, Computing in Science Engineering, 9, 90,
  \dodoi{10.1109/MCSE.2007.55}

\bibitem[{{Kingma} \& {Ba}(2014)}]{Kingma2014}
{Kingma}, D.~P., \& {Ba}, J. 2014, arXiv e-prints, arXiv:1412.6980.
\newblock \doarXiv{1412.6980}

\bibitem[{{Kochanek} {et~al.}(2017){Kochanek}, {Shappee}, {Stanek}, {Holoien},
  {Thompson}, {Prieto}, {Dong}, {Shields}, {Will}, {Britt}, {Perzanowski}, \&
  {Pojma{\'n}ski}}]{Kochanek2017}
{Kochanek}, C.~S., {Shappee}, B.~J., {Stanek}, K.~Z., {et~al.} 2017, \pasp,
  129, 104502, \dodoi{10.1088/1538-3873/aa80d9}

\bibitem[{{Lightkurve Collaboration}(2020)}]{LightkurveDocs}
{Lightkurve Collaboration}. 2020, How to remove scattered light from TESS data
  using the \texttt{RegressionCorrector}?,
  \url{https://docs.lightkurve.org/tutorials/04-how-to-remove-tess-scattered-light-using-regressioncorrector.html}

\bibitem[{{Lightkurve Collaboration} {et~al.}(2018){Lightkurve Collaboration},
  {Cardoso}, {Hedges}, {Gully-Santiago}, {Saunders}, {Cody}, {Barclay}, {Hall},
  {Sagear}, {Turtelboom}, {Zhang}, {Tzanidakis}, {Mighell}, {Coughlin}, {Bell},
  {Berta-Thompson}, {Williams}, {Dotson}, \& {Barentsen}}]{Lightkurve2018}
{Lightkurve Collaboration}, {Cardoso}, J.~V.~d.~M., {Hedges}, C., {et~al.}
  2018, {Lightkurve: Kepler and TESS time series analysis in Python},
  Astrophysics Source Code Library.
\newblock \doeprint{1812.013}

\bibitem[{Liu {et~al.}(2007)Liu, San~Liang, \& Weisberg}]{Liu2007}
Liu, Y., San~Liang, X., \& Weisberg, R.~H. 2007, Journal of Atmospheric and
  Oceanic Technology, 24, 2093, \dodoi{10.1175/2007JTECHO511.1}

\bibitem[{{Llama} {et~al.}(2012){Llama}, {Jardine}, {Mackay}, \&
  {Fares}}]{Llama2012}
{Llama}, J., {Jardine}, M., {Mackay}, D.~H., \& {Fares}, R. 2012, \mnras, 422,
  L72, \dodoi{10.1111/j.1745-3933.2012.01239.x}

\bibitem[{{Lomb}(1976)}]{Lomb1976}
{Lomb}, N.~R. 1976, \apss, 39, 447, \dodoi{10.1007/BF00648343}

\bibitem[{{Lu} {et~al.}(2020){Lu}, {Angus}, {Ag{\"u}eros}, {Blancato}, {Ness},
  {Rowland}, {Curtis}, \& {Grunblatt}}]{Lu2020}
{Lu}, Y.~L., {Angus}, R., {Ag{\"u}eros}, M.~A., {et~al.} 2020, \aj, 160, 168,
  \dodoi{10.3847/1538-3881/abada4}

\bibitem[{{Mackay} {et~al.}(2004){Mackay}, {Jardine}, {Collier Cameron},
  {Donati}, \& {Hussain}}]{Mackay2004}
{Mackay}, D.~H., {Jardine}, M., {Collier Cameron}, A., {Donati}, J.~F., \&
  {Hussain}, G.~A.~J. 2004, \mnras, 354, 737,
  \dodoi{10.1111/j.1365-2966.2004.08233.x}

\bibitem[{{Marilli} {et~al.}(2007){Marilli}, {Frasca}, {Covino}, {Alcal{\'a}},
  {Catalano}, {Fern{\'a}ndez}, {Arellano Ferro}, {Rubio-Herrera}, \&
  {Spezzi}}]{Marilli2007}
{Marilli}, E., {Frasca}, A., {Covino}, E., {et~al.} 2007, \aap, 463, 1081,
  \dodoi{10.1051/0004-6361:20066458}

\bibitem[{{Mathur} {et~al.}(2010){Mathur}, {Garc{\'\i}a}, {R{\'e}gulo},
  {Creevey}, {Ballot}, {Salabert}, {Arentoft}, {Quirion}, {Chaplin}, \&
  {Kjeldsen}}]{Mathur2010}
{Mathur}, S., {Garc{\'\i}a}, R.~A., {R{\'e}gulo}, C., {et~al.} 2010, \aap, 511,
  A46, \dodoi{10.1051/0004-6361/200913266}

\bibitem[{{McQuillan} {et~al.}(2013){McQuillan}, {Aigrain}, \&
  {Mazeh}}]{McQuillan2013}
{McQuillan}, A., {Aigrain}, S., \& {Mazeh}, T. 2013, \mnras, 432, 1203,
  \dodoi{10.1093/mnras/stt536}

\bibitem[{{McQuillan} {et~al.}(2014){McQuillan}, {Mazeh}, \&
  {Aigrain}}]{McQuillan2014}
{McQuillan}, A., {Mazeh}, T., \& {Aigrain}, S. 2014, \apjs, 211, 24,
  \dodoi{10.1088/0067-0049/211/2/24}

\bibitem[{{Netto} \& {Valio}(2020)}]{Netto2020}
{Netto}, Y., \& {Valio}, A. 2020, \aap, 635, A78,
  \dodoi{10.1051/0004-6361/201936219}

\bibitem[{{Newton} {et~al.}(2018){Newton}, {Mondrik}, {Irwin}, {Winters}, \&
  {Charbonneau}}]{Newton2018}
{Newton}, E.~R., {Mondrik}, N., {Irwin}, J., {Winters}, J.~G., \&
  {Charbonneau}, D. 2018, \aj, 156, 217, \dodoi{10.3847/1538-3881/aad73b}

\bibitem[{{Nielsen} {et~al.}(2019){Nielsen}, {Gizon}, {Cameron}, \&
  {Miesch}}]{Nielsen2019}
{Nielsen}, M.~B., {Gizon}, L., {Cameron}, R.~H., \& {Miesch}, M. 2019, \aap,
  622, A85, \dodoi{10.1051/0004-6361/201834373}

\bibitem[{{Oelkers} \& {Stassun}(2018)}]{Oelkers2018b}
{Oelkers}, R.~J., \& {Stassun}, K.~G. 2018, \aj, 156, 132,
  \dodoi{10.3847/1538-3881/aad68e}

\bibitem[{{Oelkers} {et~al.}(2018){Oelkers}, {Rodriguez}, {Stassun}, {Pepper},
  {Somers}, {Kafka}, {Stevens}, {Beatty}, {Siverd}, {Lund}, {Kuhn}, {James}, \&
  {Gaudi}}]{Oelkers2018a}
{Oelkers}, R.~J., {Rodriguez}, J.~E., {Stassun}, K.~G., {et~al.} 2018, \aj,
  155, 39, \dodoi{10.3847/1538-3881/aa9bf4}

\bibitem[{Paszke {et~al.}(2019)Paszke, Gross, Massa, Lerer, Bradbury, Chanan,
  Killeen, Lin, Gimelshein, Antiga, Desmaison, Kopf, Yang, DeVito, Raison,
  Tejani, Chilamkurthy, Steiner, Fang, Bai, \& Chintala}]{Pytorch2019}
Paszke, A., Gross, S., Massa, F., {et~al.} 2019, in Advances in Neural
  Information Processing Systems 32, ed. H.~Wallach, H.~Larochelle,
  A.~Beygelzimer, F.~d'Alch\'{e} Buc, E.~Fox, \& R.~Garnett (Curran Associates,
  Inc.), 8024--8035.
\newblock
  \url{http://papers.neurips.cc/paper/9015-pytorch-an-imperative-style-high-performance-deep-learning-library.pdf}

\bibitem[{{Pepper} {et~al.}(2007){Pepper}, {Pogge}, {DePoy}, {Marshall},
  {Stanek}, {Stutz}, {Poindexter}, {Siverd}, {O'Brien}, {Trueblood}, \&
  {Trueblood}}]{Pepper2007}
{Pepper}, J., {Pogge}, R.~W., {DePoy}, D.~L., {et~al.} 2007, \pasp, 119, 923,
  \dodoi{10.1086/521836}

\bibitem[{{Perez} \& {Granger}(2007)}]{iPython2007}
{Perez}, F., \& {Granger}, B.~E. 2007, Computing in Science Engineering, 9, 21,
  \dodoi{10.1109/MCSE.2007.53}

\bibitem[{{Reinhold} \& {Hekker}(2020)}]{Reinhold2020}
{Reinhold}, T., \& {Hekker}, S. 2020, \aap, 635, A43,
  \dodoi{10.1051/0004-6361/201936887}

\bibitem[{{Ricker} {et~al.}(2015){Ricker}, {Winn}, {Vanderspek}, {Latham},
  {Bakos}, {Bean}, {Berta-Thompson}, {Brown}, {Buchhave}, {Butler}, {Butler},
  {Chaplin}, {Charbonneau}, {Christensen-Dalsgaard}, {Clampin}, {Deming},
  {Doty}, {De Lee}, {Dressing}, {Dunham}, {Endl}, {Fressin}, {Ge}, {Henning},
  {Holman}, {Howard}, {Ida}, {Jenkins}, {Jernigan}, {Johnson}, {Kaltenegger},
  {Kawai}, {Kjeldsen}, {Laughlin}, {Levine}, {Lin}, {Lissauer}, {MacQueen},
  {Marcy}, {McCullough}, {Morton}, {Narita}, {Paegert}, {Palle}, {Pepe},
  {Pepper}, {Quirrenbach}, {Rinehart}, {Sasselov}, {Sato}, {Seager},
  {Sozzetti}, {Stassun}, {Sullivan}, {Szentgyorgyi}, {Torres}, {Udry}, \&
  {Villasenor}}]{Ricker2020}
{Ricker}, G.~R., {Winn}, J.~N., {Vanderspek}, R., {et~al.} 2015, Journal of
  Astronomical Telescopes, Instruments, and Systems, 1, 014003,
  \dodoi{10.1117/1.JATIS.1.1.014003}

\bibitem[{{R{\"u}diger} {et~al.}(2019){R{\"u}diger}, {K{\"u}ker},
  {K{\"a}pyl{\"a}}, \& {Strassmeier}}]{Rudiger2019}
{R{\"u}diger}, G., {K{\"u}ker}, M., {K{\"a}pyl{\"a}}, P.~J., \& {Strassmeier},
  K.~G. 2019, \aap, 630, A109, \dodoi{10.1051/0004-6361/201935280}

\bibitem[{{Santos} {et~al.}(2019){Santos}, {Garc{\'\i}a}, {Mathur}, {Bugnet},
  {van Saders}, {Metcalfe}, {Simonian}, \& {Pinsonneault}}]{Santos2019}
{Santos}, A.~R.~G., {Garc{\'\i}a}, R.~A., {Mathur}, S., {et~al.} 2019, \apjs,
  244, 21, \dodoi{10.3847/1538-4365/ab3b56}

\bibitem[{{Scargle}(1982)}]{Scargle1982}
{Scargle}, J.~D. 1982, \apj, 263, 835, \dodoi{10.1086/160554}

\bibitem[{{Schrijver} \& {Harvey}(1994)}]{Schrijver1994}
{Schrijver}, C.~J., \& {Harvey}, K.~L. 1994, \solphys, 150, 1,
  \dodoi{10.1007/BF00712873}

\bibitem[{{Shappee} {et~al.}(2014){Shappee}, {Prieto}, {Grupe}, {Kochanek},
  {Stanek}, {De Rosa}, {Mathur}, {Zu}, {Peterson}, {Pogge}, {Komossa}, {Im},
  {Jencson}, {Holoien}, {Basu}, {Beacom}, {Szczygie{\l}}, {Brimacombe},
  {Adams}, {Campillay}, {Choi}, {Contreras}, {Dietrich}, {Dubberley},
  {Elphick}, {Foale}, {Giustini}, {Gonzalez}, {Hawkins}, {Howell}, {Hsiao},
  {Koss}, {Leighly}, {Morrell}, {Mudd}, {Mullins}, {Nugent}, {Parrent},
  {Phillips}, {Pojmanski}, {Rosing}, {Ross}, {Sand}, {Terndrup}, {Valenti},
  {Walker}, \& {Yoon}}]{Shappee2014}
{Shappee}, B.~J., {Prieto}, J.~L., {Grupe}, D., {et~al.} 2014, \apj, 788, 48,
  \dodoi{10.1088/0004-637X/788/1/48}

\bibitem[{{Thomas} {et~al.}(2019){Thomas}, {Chaplin}, {Davies}, {Howe},
  {Santos}, {Elsworth}, {Miglio}, {Campante}, \& {Cunha}}]{Thomas2019}
{Thomas}, A. E.~L., {Chaplin}, W.~J., {Davies}, G.~R., {et~al.} 2019, \mnras,
  485, 3857, \dodoi{10.1093/mnras/stz672}

\bibitem[{Torrence \& Compo(1998)}]{Torrence1998}
Torrence, C., \& Compo, G. 1998, A practical guide to wavelet analysis, B. Am.
  Meteorol. Soc., 79, 61--78

\bibitem[{{van Ballegooijen} {et~al.}(1998){van Ballegooijen}, {Cartledge}, \&
  {Priest}}]{vanBallegooijen1998}
{van Ballegooijen}, A.~A., {Cartledge}, N.~P., \& {Priest}, E.~R. 1998, \apj,
  501, 866, \dodoi{10.1086/305823}

\bibitem[{{van Saders} {et~al.}(2016){van Saders}, {Ceillier}, {Metcalfe},
  {Silva Aguirre}, {Pinsonneault}, {Garc{\'\i}a}, {Mathur}, \&
  {Davies}}]{vanSaders2016}
{van Saders}, J.~L., {Ceillier}, T., {Metcalfe}, T.~S., {et~al.} 2016, \nat,
  529, 181, \dodoi{10.1038/nature16168}

\bibitem[{{van Saders} {et~al.}(2019){van Saders}, {Pinsonneault}, \&
  {Barbieri}}]{vanSaders2019}
{van Saders}, J.~L., {Pinsonneault}, M.~H., \& {Barbieri}, M. 2019, \apj, 872,
  128, \dodoi{10.3847/1538-4357/aafafe}

\bibitem[{{Vanderspek} {et~al.}(2018){Vanderspek}, {Doty}, {Fausnaugh},
  {Villase{\~n}or}, {Jenkins}, {Berta-Thompson}, {Burke}, \&
  {Ricker}}]{TESSHandbook}
{Vanderspek}, R., {Doty}, J.~P., {Fausnaugh}, M., {et~al.} 2018, TESS
  Instrument Handbook,
  \url{https://archive.stsci.edu/missions/tess/doc/TESS_Instrument_Handbook_v0.1.pdf}

\bibitem[{{Virtanen} {et~al.}(2020){Virtanen}, {Gommers}, {Oliphant},
  {Haberland}, {Reddy}, {Cournapeau}, {Burovski}, {Peterson}, {Weckesser},
  {Bright}, {van der Walt}, {Brett}, {Wilson}, {Jarrod Millman}, {Mayorov},
  {Nelson}, {Jones}, {Kern}, {Larson}, {Carey}, {Polat}, {Feng}, {Moore}, {Vand
  erPlas}, {Laxalde}, {Perktold}, {Cimrman}, {Henriksen}, {Quintero}, {Harris},
  {Archibald}, {Ribeiro}, {Pedregosa}, {van Mulbregt}, \&
  {Contributors}}]{Scipy2020}
{Virtanen}, P., {Gommers}, R., {Oliphant}, T.~E., {et~al.} 2020, Nature
  Methods, 17, 261, \dodoi{https://doi.org/10.1038/s41592-019-0686-2}

\bibitem[{{W}es {M}c{K}inney(2010)}]{Pandas2010}
{W}es {M}c{K}inney. 2010, in {P}roceedings of the 9th {P}ython in {S}cience
  {C}onference, ed. {S}t\'efan van~der {W}alt \& {J}arrod {M}illman, 56 -- 61

\end{thebibliography}
\bibliographystyle{aasjournal}

\end{document}